\documentclass[conference, a4paper]{IEEEtran}

\IEEEoverridecommandlockouts
\usepackage{graphicx,color,float}
\usepackage{amssymb,amsmath,times,wasysym,comment}
\usepackage{cite}
\usepackage{epsfig}
\UseRawInputEncoding
\usepackage{makecell}
\usepackage{ctable}
\usepackage{tabularx,booktabs}
\usepackage{diagbox}
\newcolumntype{C}{>{\centering\arraybackslash}X}
\usepackage[hyphens]{url}
\usepackage{hyperref}
\usepackage[acronym]{glossaries}
\hypersetup{breaklinks=true}

\title{A Survey on Non-Geostationary Satellite Systems: The Communication Perspective}

 \author{\normalsize Hayder Al-Hraishawi,~\IEEEmembership{Member,~IEEE}, Houcine Chougrani,  Steven Kisseleff, ~\IEEEmembership{Member,~IEEE},\\ Eva Lagunas, ~\IEEEmembership{Senior Member,~IEEE}, and Symeon Chatzinotas, ~\IEEEmembership{Senior Member,~IEEE}\\\vspace{-5mm}

\thanks{The authors are with the Interdisciplinary Centre for Security, Reliability and Trust (SnT), University of Luxembourg, Luxembourg.
Corresponding author: \textit{Hayder Al-Hraishawi (hayder.al-hraishawi@uni.lu)}.}

 \thanks{This research was funded in whole by the Luxembourg National Research Fund (FNR) in the frameworks of the FNR-CORE project "MegaLEO: Self-Organised Lower Earth Orbit Mega-Constellations"  (Grant no. C20/IS/14767486). For the purpose of open access, the authors have applied a Creative Commons Attribution 4.0 International (CC BY 4.0) license to any Author Accepted Manuscript version arising from this submission.}
} 

\begin{document}
	\maketitle
	\thispagestyle{plain}
    \pagestyle{plain}	
   
\begin{abstract}
	The next phase of satellite technology is being characterized by a new evolution in non-geostationary orbit (NGSO) satellites, which conveys exciting new communication capabilities to provide non-terrestrial connectivity solutions and to support a wide range of digital technologies from various industries.
    NGSO communication systems are known for a number of key features such as lower propagation delay, smaller size, and lower signal losses in comparison to the conventional geostationary orbit (GSO) satellites, which can potentially enable latency-critical applications to be provided through satellites. NGSO promises a substantial boost in communication speed and energy efficiency, and thus, tackling the main inhibiting factors of commercializing GSO satellites for broader utilization. The promised improvements of NGSO systems have motivated this paper to provide a comprehensive survey of the state-of-the-art NGSO research focusing on the communication prospects, including physical layer and radio access technologies along with the networking aspects and the overall system features and architectures. Beyond this, there are still many NGSO deployment challenges to be addressed to ensure seamless integration not only with GSO systems but also with terrestrial networks. These unprecedented challenges are also discussed in this paper, including coexistence with GSO systems in terms of spectrum access and regulatory issues, satellite constellation and architecture designs, resource management problems, and user equipment requirements. Finally, we  outline a set of innovative research directions and new opportunities for future NGSO research. 
\end{abstract}


\section{Introduction}\label{sec:intro}
Satellites have a distinctive ability of covering wide geographical areas through a minimum amount of infrastructure on the ground,  which qualifies them as an appealing solution to fulfill the growing number of diverse applications and services  either as a stand-alone system, or as an integrated satellite-terrestrial network \cite{Perez2019}. Currently, the field of satellite communications is drawing an increased attention in the global telecommunications market as several network operators start using satellites in backhauling infrastructures for connectivity and for fifth-generation (5G) system integration \cite{Giambene2018}. Recently, due to the swift rise of  “NewSpace” industries \cite{Venezia2019} that are developing small satellites with new low-cost launchers \cite{schmierer2019}, a large number of satellite operators have already planned to launch thousands of non-geostationary (NGSO) satellites to satisfy the burgeoning demand for global broadband, high-speed, ultra-reliable and low latency communications.

Furthermore, satellite systems have been contributing to deliver telecommunication services in a wide range of sectors such as aeronautical, maritime, military, rescue and disaster relief. Beyond this, NGSO systems are envisaged to be an efficient solution for future non-terrestrial networks (NTN) to meet the demanding sixth-generation (6G) system requirements in terms of both large throughput and global connectivity \cite{Giordani2021}. In this direction, the third generation partnership project (3GPP) standards group has been codifying the use of satellite systems to integrate space-airborne-terrestrial communication networks in order to support future wireless ecosystems \cite{3GPP38821v16, 3GPP38811v15}. Moreover, by harnessing satellite’s geographical independence, wireless connectivity can be extended to the underserved and unserved areas, where NGSO systems can facilitate the deployment of 5G and beyond networks. Thus, NGSO satellites are expected to play a crucial role in bridging the digital divide by  extending backhaul for 5G services and providing high-bandwidth links directly to the end users \cite{Khalil2019}.

\subsection{Background}
Geostationary orbit (GSO) satellites are orbiting at the equatorial plane at an altitude of 35,678 km with an almost zero-inclination angle. Notwithstanding GSO large coverage, these satellites cannot cover the high-latitude areas. Additionally, the communication links between GSO satellites and ground stations are vulnerable to high propagation losses, and hence, large antennas with higher transmit power are necessary \cite{Kolawole2017}. Moreover, the propagation delay of GSO satellites is high due to the long propagation path, which makes them unfavourable for delay-sensitive services. Whereas, NGSO satellites on a geocentric orbit include the low Earth orbit (LEO), medium Earth orbit (MEO), and highly elliptical orbit (HEO) satellites, which are orbiting constantly at lower altitudes than GSO satellites, and thus, their link losses and latency due to signal propagation are lower \cite{ITU2003}. Since these lower orbit satellites serve smaller coverage areas than GSO ones, a constellation of NGSO satellites is needed to achieve full Earth coverage including the high-latitude regions \cite{Wood2003}.

\begin{figure}[t!]\centering
	\includegraphics[width=0.5\textwidth]{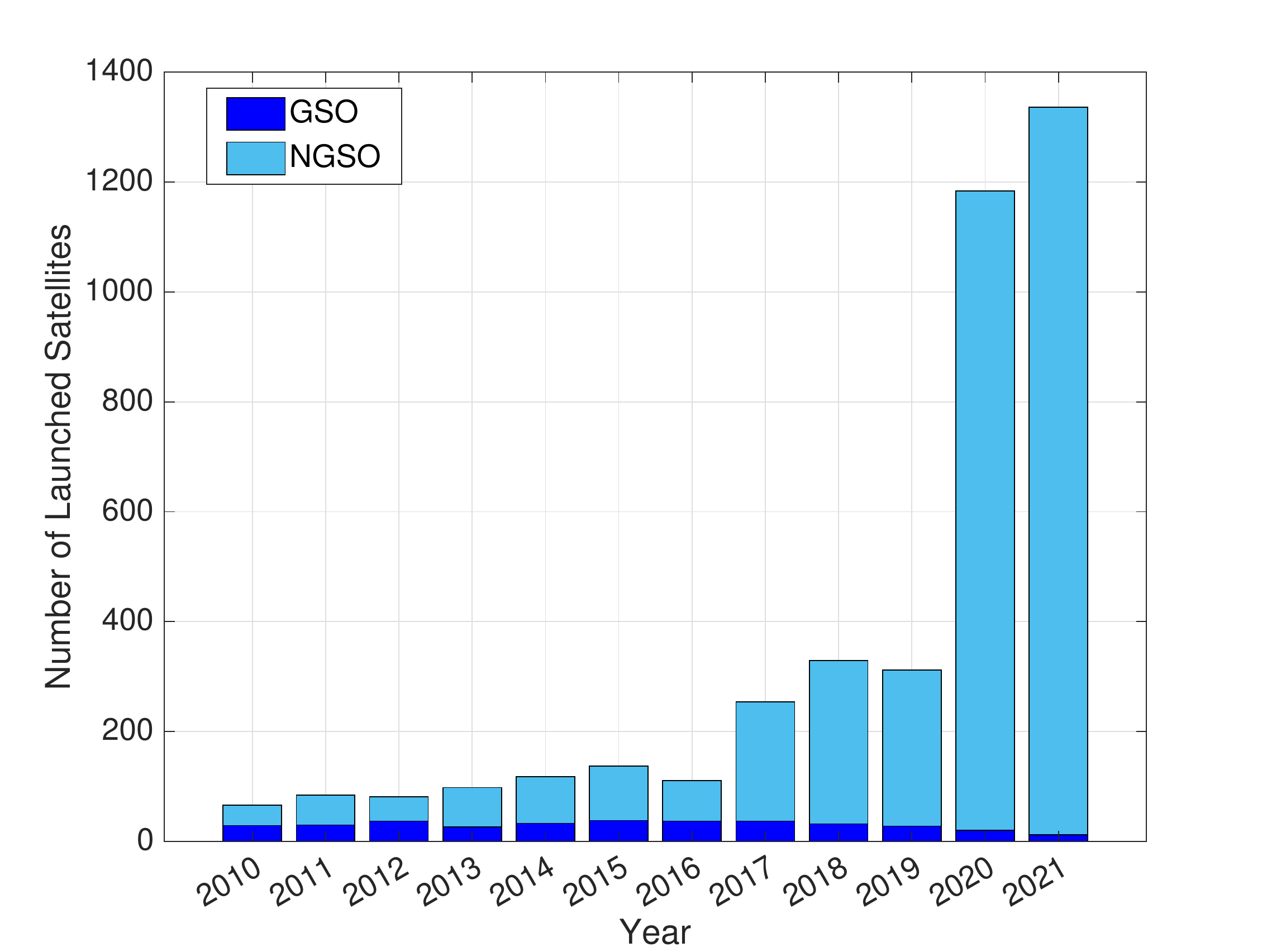}
	\caption{Comparison between GSO and NGSO in terms of the number of launched satellites per year \cite{UCS_DS}.} \vspace{-2mm}
	\label{fig:sat_count}\vspace{-2mm}
\end{figure}

All of the aforementioned NGSO advantages revive the notion of utilizing large fleets of lower orbit satellites to provide reliable, low-latency, and high-speed Internet from space, which has re-gained popularity and experienced a tremendous growth in the last few years \cite{Saad2020}. This trend is rather surprising given the unfortunate faring of past NGSO constellations, but it appears that both technological and business momentums are favorable with impressive achievements from SpaceX, SES O3B, and OneWeb \cite{Bhattacherjee2018}. In fact, between 2014 and 2016, a new wave of proposals for large LEO constellations emerged with the target of providing global broadband services \cite{Portillo2019}. 
Specifically, the number of satellites were launched into space has dramatically increased according to the recent satellite database released by the Union of Concerned Scientists (UCS) \cite{UCS_DS}. This database has  listed  more than 4,000 operational satellites currently in orbit around Earth with huge difference between the number of GSO and NGSO satellites in favor of the latter as depicted in Fig. \ref{fig:sat_count}. 

Furthermore, the most recent developments in NGSO systems empower satellites to manage narrow steerable beams covering a relatively broad area, which facilities the use of smaller and lower cost equipment at the user terminals \cite{Guan2019}. Hence, NGSO satellite capabilities of ubiquitous coverage and connectivity can be leveraged for provisioning resiliency and continuity of 5G services to the mobile platforms such as on-board aircraft, high-speed trains, sea-going vessels, and land-based vehicles that are beyond the reach of a cell site \cite{Di2019a}. More importantly, the offered capacities by NGSO satellites can be further increased by utilizing high frequencies along with throughput enhancement techniques such as spectrum sharing, cooperative gateway diversity, user clustering and interference mitigation, and multiple antenna communications  \cite{Hayder2021a}. For instance, the emerging NGSO satellites and mega constellations such as SES O3b, OneWeb, Telesat, and Starlink have a system capacity reaching the terabits-per-second level \cite{Su2019}.

In addition to the NGSO unique capabilities in providing global coverage, low-latency communication, and high-speed Internet access points, these systems can ameliorate the way satellite missions are designed and operated in the near future \cite{Babich2020}. In particular, the recent technological progress has evolved the possibility of constructing a chain production of cheaper NGSO satellites with very short lifespans \cite{Singh2020}. Accordingly, the satellite infrastructure will be more regularly upgraded, and thus, the payload design can be more innovative in terms of on-board technologies \cite{Stock2020}. Evidently, NGSO satellites can create new capabilities and services for different enterprise verticals and could also open up many new opportunities for innovative applications  \cite{Latio2021}. However, that comes with some important questions about their operations and the required developments. Thereby, the purpose of this work is providing a survey of key research progress in this rising field from the communication perspective, identifying the key deployment challenges, along with highlighting promising future research directions for NGSO systems.

% Section II. Review of existing surveys
%========================================
\subsection{Prior Related Surveys}
%========================================
Over the last few years, a number of good surveys and tutorials pertained to satellite communications appeared in the literature, \cite{Niephaus2016,Radhakrishnan2016,Liu2018a,Zeed2019,Burleigh2019,Kodheli2020,Li2020,Mayorga2020,Saeed2020,Madni2020,Guidotti2020,Rinaldi2020,Wang2020a,Fang2021}, to report and study the technical developments and challenges, including satellite network architectures, attributes and applications of lower orbit satellites, satellite-terrestrial systems integration, and small satellite systems. In the following, research scope and contributions of the relevant surveys will be briefly presented. Afterwards, a comparison between these surveys and our work in this paper will be summarized at a glance in Table \ref{table:surveys_comparison} in order to point out the distinctive contribution of our survey. 

Convergence of satellite and terrestrial networks is surveyed in \cite{Niephaus2016} with focusing on scenarios in which satellite networks complement existing terrestrial infrastructures. In this, the technical challenges associated with the convergence of satellite and terrestrial networks to provide ubiquitous connectivity in rural and remote areas are identified. The work in \cite{Radhakrishnan2016} surveys the research efforts for implementing inter-satellite communication for small satellite systems, by reviewing various constellation design parameters  within the first three layers of OSI model, i.e., physical, data link, and network layer. The available research works on space-air-ground integrated networks have been surveyed in \cite{Liu2018a}, where the aspects of network design, resource allocation and optimization, protocol planning, and performance analysis are covered. The work in \cite{Liu2018a} has also pointed out the key technical challenges and design issues for deploying space-air-ground integrated networks and provided some future research directions that might be worthy of further investigations.

Moreover, the limitations of land mobile satellite (LMS) systems in terms of connectivity, stability, and reliability are studied in \cite{Zeed2019}, where the LMS is considered as a satellite-based communication system that can serve ground users in different areas. LMS systems are overviewed based on satellite orbits, operating frequency bands, and signal propagation along with highlighting some future research challenges. Besides, the recent advances and development trends in the field of small satellites are explored in \cite{Burleigh2019} with emphasizing the aspects of satellite communications such as the use of higher frequency bands, optical communications, new protocols, and the applicable architectures and use cases. 

The survey in \cite{Kodheli2020} has captured the recent technical advances in scientific, industrial and standardization analyses in the domain of satellite communications with presenting the important research directions for satellite communication applications and use cases such as new constellation types, on-board processing capabilities, non-terrestrial networks and space-based data collection and processing. 
A review of the state-of-the-art research progress of satellite communications covering LMS communication networks, hybrid satellite-terrestrial relay networks, and satellite-terrestrial integrated networks is provided in \cite{Li2020} under the framework of physical-layer security. The potentials and challenges of satellite-based Internet of things (IoT) architecture have been also studied  in \cite{Li2020}, along with popularized performance metrics in order to evaluate system security.

%[small satellite systems]
Authors in \cite {Mayorga2020} have reviewed the connectivity challenges in LEO small-satellite constellations, along with the essential architectural and technological components that will enable 5G connectivity through LEO satellites. Reference \cite{Saeed2020} reviews the literature of CubeSat communications through exploring some relevant aspects such as channel modeling, modulation and coding, coverage, networking, and constellation-and-coverage issues, along with highlighting future research challenges for enabling the concept of Internet of space things. 
Networking and routing aspects of small satellite systems are considered in the survey in \cite{Madni2020} with special focus on inter-satellite routing protocols and the performance of delay tolerant (DTN) and non-delay tolerant (Non-DTN) schemes under different CubeSat network sittings.

%[NTN]
In \cite{Guidotti2020}, the architectural and technological challenges of integrating satellites into 5G systems for both physical and medium access control (MAC) layers has been discussed in the context of the proposed 3GPP NTN systems. In this, different NTN scenarios for satellite-based 5G communications have been analyzed and reviewed in terms of satellite orbits, payload types, protocol design, and radio interfaces. Similarly, the work in \cite{Rinaldi2020} studies the 3GPP NTN features and their deployment potentials  within 5G and beyond networks through reviewing current 3GPP research activities, discussing the open issues of NTN over the wireless communication landscape, and identifying future research directions of NTN evolution in connection to terrestrial communications.

Similarly, the requirements of satellite-terrestrial network convergence are reviewed in \cite{Wang2020a} with summarizing the relevant architectures of existing literature, classifying the taxonomy of researches on satellite-terrestrial networks, and presenting the performance evaluation works in different satellite-terrestrial networks, together with providing the state-of-the-art of standardization, projects and the key application areas of satellite-terrestrial networks.
The work in \cite{Fang2021} has studied the challenges of deploying hybrid satellite-terrestrial networks and explored the complicated coupling relationships therein. In \cite{Fang2021}, the setup of hybrid satellite-terrestrial networks is considered as a combination of basic cooperative models that contain the main entities of satellite-terrestrial integration and  are simpler and tractable compared to  the large-scale heterogeneous hybrid satellite-terrestrial networks.

The abovementioned surveys have addressed important aspects of satellite developments but there still lacks a survey providing comprehensive discussions on the whole  multi-orbit NGSO communication system aspects, presenting NGSO integration challenges within the existing wireless networks, and identifying future research directions and opportunities. This observation has motivated composing this article to provide an in-depth discussion on the communication aspects of NGSO satellites with current and future terrestrial networks to ensure full coverage consistent with the existing satellite constellations and GSO systems. In addition, regarding NGSO challenges, the existing survey articles provide only high-level discussions. For instance, the regulatory and coexistence challenges have been briefly covered in the previous works, while the user equipment requirements and advances have not been explored in the open literature. Furthermore, it is essential to have a wide-ranging  survey as NGSO systems have started to gain momentum recently in both academia and industry, accordingly such a survey can benefit readers from both communities. Therefore, this survey paper aims at exploring the state-of-the-art NGSO research findings from the communication perspective, discussing the NGSO deployment hurdles, and providing future opportunities for further NGSO research activities.

\begin{table*}[!t]
	\centering
	\caption{Comparison with previous surveys}
	\label{tab:comparison}
	\begin{tabularx}{\textwidth}{@{}l*{15}{C}c@{}}
		\midrule \label{table:surveys_comparison}
		\hspace{2.1mm}\diagbox[width=15em]{\shortstack{Covered\\scope}}{Reference}  &  
		\rotatebox[origin=c]{0}{\shortstack{\cite{Niephaus2016}\\2016}}&
		\rotatebox[origin=c]{0}{\shortstack {{\cite{Radhakrishnan2016}} \\2016}}&
		\rotatebox[origin=c]{0}{\shortstack{\cite{Liu2018a}\\2018}}&
		\rotatebox[origin=c]{0}{\shortstack{\cite{Zeed2019}\\2019}} & 
		\rotatebox[origin=c]{0}{\shortstack{\cite{Burleigh2019}\\2019}}&
		\rotatebox[origin=c]{0}{\shortstack {\cite{Kodheli2020} \\ 2020}}  & \rotatebox[origin=c]{0}{\shortstack {{\cite{Li2020}}\\2020}} &
		 \rotatebox[origin=c]{0}{\shortstack{\cite{Mayorga2020}\\2020}} & \rotatebox[origin=c]{0}{\shortstack{\cite{Saeed2020}\\2020}}& 
		\rotatebox[origin=c]{0}{\shortstack{\cite{Madni2020}\\2020}}&
		\rotatebox[origin=c]{0}{\shortstack{\cite{Guidotti2020}\\2020}}&
		\rotatebox[origin=c]{0}{\shortstack{\cite{Rinaldi2020}\\2020}}&    \rotatebox[origin=c]{0}{\shortstack {\cite{Wang2020a} \\ 2020}}& \rotatebox[origin=c]{0}{\shortstack{\cite{Fang2021} \\ 2021}}&  \rotatebox[origin=c]{0}{\shortstack{Our \\ paper\\2022}}  \\ \midrule
	Space-based Internet systems &   & \checkmark & \checkmark & \checkmark & \checkmark & \checkmark &   & \checkmark & \checkmark & \checkmark &   &   &   &   & \checkmark \\
 		NGSO space missions                     &   & \checkmark & \checkmark & \checkmark & \checkmark & \checkmark & \checkmark & \checkmark & \checkmark & \checkmark &   &   &   &   & \checkmark \\
 		Regulatory and coexistence issues       &   &   &   &   &   &   &   &   &   &   &   &   &   &   & \checkmark \\
 		Constellation design methods            &   & \checkmark & \checkmark & \checkmark & \checkmark & \checkmark & \checkmark & \checkmark & \checkmark & \checkmark & \checkmark & \checkmark &   &   & \checkmark \\
 		NGSO operational challenges             & \checkmark & \checkmark & \checkmark & \checkmark & \checkmark & \checkmark &   & \checkmark & \checkmark & \checkmark & \checkmark & \checkmark & \checkmark & \checkmark & \checkmark \\
 		User equipment requirements             &   &   &   &   &   &   &   &   &   &   &   &   &   &   & \checkmark \\
 		Inter-satellite connectivity            &   & \checkmark & \checkmark & \checkmark &   &   & \checkmark &   & \checkmark & \checkmark & \checkmark & \checkmark & \checkmark & \checkmark & \checkmark \\
 		NGSO active antenna systems             &   &   &   &   &   & \checkmark &   &   &   &   &   &   &   &   & \checkmark \\
 		Waveform design and access schemes      & \checkmark & \checkmark & \checkmark & \checkmark & \checkmark & \checkmark & \checkmark & \checkmark & \checkmark &   & \checkmark & \checkmark & \checkmark &   & \checkmark \\
 		Software-defined satellites             & \checkmark &   & \checkmark &   & \checkmark & \checkmark & \checkmark &   & \checkmark &   & \checkmark & \checkmark & \checkmark & \checkmark & \checkmark \\
 		In-space backhauling                    &   &   & \checkmark &   &   & \checkmark &   & \checkmark & \checkmark &   & \checkmark & \checkmark & \checkmark & \checkmark & \checkmark \\
 		Satellite network slicing               &   &   &   &   &   & \checkmark &   & \checkmark &   &   & \checkmark & \checkmark & \checkmark &   & \checkmark \\
 		Resource optimization                   & \checkmark & \checkmark & \checkmark &   & \checkmark & \checkmark &   & \checkmark & \checkmark & \checkmark & \checkmark & \checkmark & \checkmark & \checkmark & \checkmark \\
 		Interference management                 &   & \checkmark & \checkmark &   &   & \checkmark & \checkmark & \checkmark &   &   &   & \checkmark & \checkmark & \checkmark & \checkmark \\
 		Secure communications                   &   & \checkmark & \checkmark &   & \checkmark &   & \checkmark &   &   &   &   & \checkmark & \checkmark & \checkmark & \checkmark \\
 		Space broadband connectivity            &   &   &   &   &   &   &   &   &   &   &   &   &   &   & \checkmark \\
 		Open RAN architecture                   &   &   &   &   &   &   &   &   &   &   &   &   &   &   & \checkmark \\
		\bottomrule
	\end{tabularx}
\end{table*}

\subsection{Scope and Contributions}
The major objective of this paper is to give the reader the technological trends and future prospects of the multi-orbit NGSO satellite communication systems including space-based Internet providers and the small satellites for space downstream missions. This paper differs from the existing surveys on  satellite communications in the following aspects. First, we present a comprehensive survey on the NGSO communication system aspects starting from the physical layer up to the application layer and the overall structural design visions, which is the central theme of the this paper. In addition, this survey summarizes NGSO satellite features and use cases to provide a quick reference for both researchers and practitioners.	Next, we provide a wide-ranging analysis for NGSO system development, deployment, and integration challenges, as well as the operational issues, for which potential solutions are also provided. Further, several innovative visions and future research directions motivated by utilizing NGSO systems are discussed in the context of other 5G technologies.

In a nutshell, the key contributions of this paper can be summarized as follows:
\begin{itemize}
	\item A detailed review and classification for the different NGSO systems are presented based on their applications. Specifically, both the emerging NGSO mega-constellation for broadband services and the space downstream missions are discussed.
	
	\item An in-depth discussion on the NGSO communication systems is provided by exploring the physical layer technologies and radio access schemes along with exploring the networking aspects, and the overall system characteristics and architectures.
		
	\item A state-of-the-art knowledge and studies are discussed regarding  NGSO satellite deployment challenges including coexistence with GSO systems and regulatory issues, satellite constellation designs, system operational issues, and user hardware capabilities and requirements.

	\item The expected evolution in satellite and terrestrial-satellite integrated communication systems are extensively studied alongside with the relevant innovative research directions of utilizing NGSO features for versatile communication infrastructure systems.
	
	\item New application scenarios of NGSO satellites are presented with exploring the potential technical advances in the future communication systems and networking due to NGSO involvements.  
\end{itemize} 

This paper can serve as a valuable resource for understanding the current research contributions in this evolving area of satellite communications that may probably initiate further research efforts.

\subsection{Structure and Organization}
\begin{figure*}
	\centering
	\setlength\fboxsep{0pt}
	\setlength\fboxrule{0.25pt}
	\includegraphics[width=0.92\textwidth]{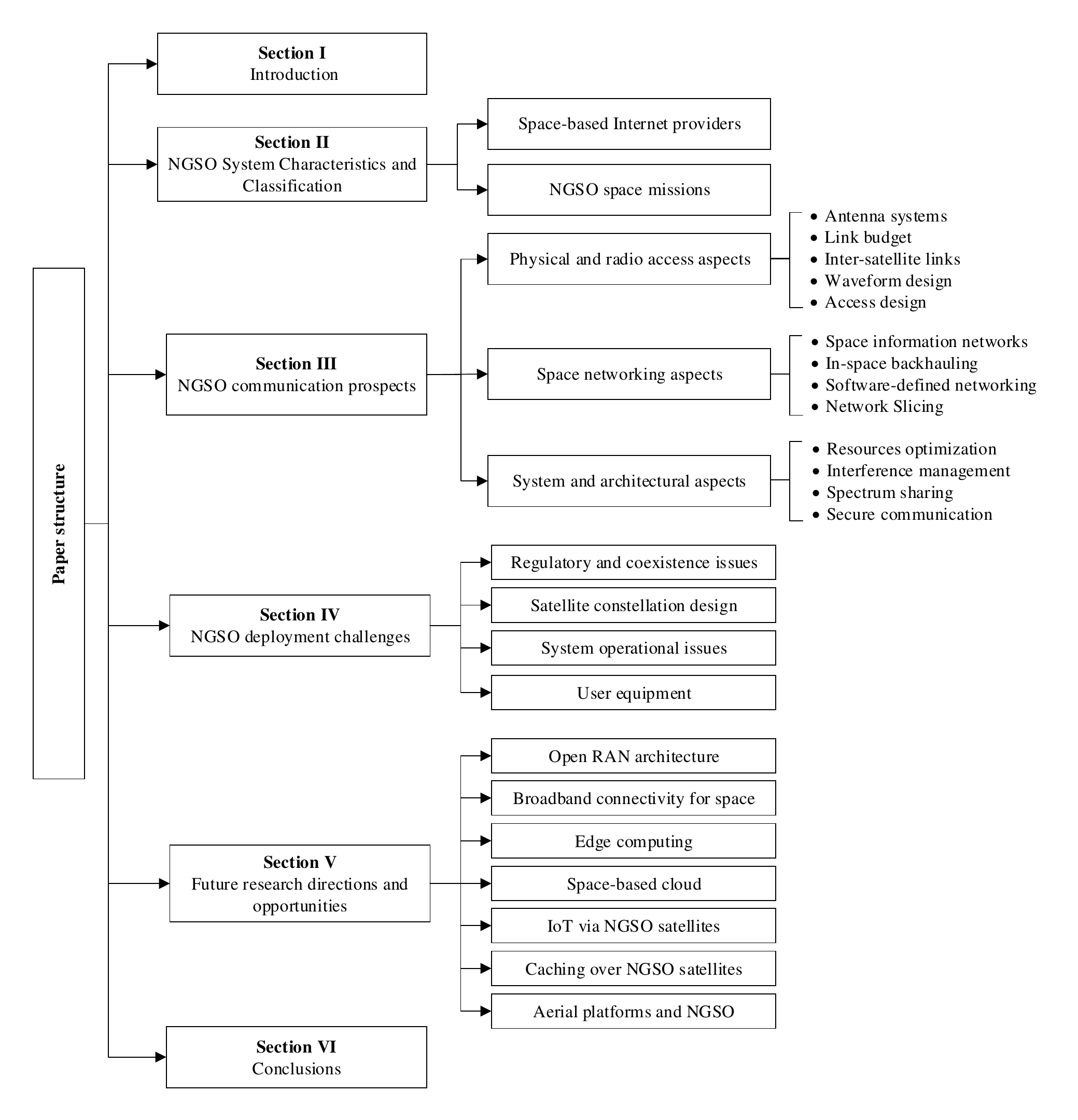} \vspace{-1mm}
	\caption{Structure and organization of the paper}
	\label{fig:paper_structure}
\end{figure*}

The reminder of this paper is organized as follows. NGSO system definition and classification are elaborated in Section \ref{sec:ngso_classification}. In Section \ref{sec:ngso_communication}, NGSO communication prospects are discussed starting from the physical layer technologies and radio access schemes along with exploring the networking aspects, and then, the overall system characteristics and architectures are studied. Section \ref{sec:ngso_challenges} presents the NGSO deployment challenges that require more research efforts for enabling seamless integration and efficient operations. Future research directions and opportunities are described in Section \ref{sec:future_research}. This
article is then concluded in Section \ref{sec:conclusions}.
For the sake of clarity, we provide Fig. \ref{fig:paper_structure} to show the structure and organization of this paper, we also list the acronyms that will be frequently used in this paper along with their definitions in Table \ref{tab:acronyms} for ease of reference.

\begin{table*}[!t]
\caption{List of Important Acronyms} \label{tab:acronyms}
\begin{tabularx}{\textwidth}{l l|l l}
\hline
\textbf{Acronym} & \textbf{Definition}                           & \textbf{Acronym} & \textbf{Definition} \\ \hline
5G               & Fifth-Generation                              & LEO          & Low Earth Orbit   \\ 
6G               & Sixth-Generation                              & LMS              & Land Mobile Satellite  \\ 
AAS              & American Astronomical Society                 & LTE-A            & Long Term Evolution-Advance   \\ 
ADS-B            & Automatic Dependent Surveillance – Broadcast  & MAC              & Medium Access   Control   \\ 
AI               & Artificial Intelligence                       & MEO              & Medium Earth Orbit   \\ 
AIS              & Automatic Identification System               & ML               & Machine Learning   \\
AoA              & Angle-of-Arrival                              & mmWave           & millimeter Wave  \\ 
BCT              & Block-Chain Technology                        & MSS              & Mobile Satellite Service  \\ 
CA               & Carrier Aggregation                           & NASA             & National Aeronautics and   Space Administration (USA) \\ 
CAPEX            & Capital Expenditures                          & NB               & Narrow-Band        \\ 
CCSDS            & Consultative Committee for Space Data Systems & NCC              & Network Control Centre \\ 
CDMA             & Code Division Multiple Access                 & NFV              & Network Function Virtualization   \\ 
CNR              & Carrier-to-Noise Ratio                        & NGSO             & Non-Geostationary Orbit  \\ 
CSMA             & Carrier Sense Multiple Access                 & NMC              & Network Management Centre    \\ 
DDoS             & Distributed Denial-of-Service                 & NOMA             & Non-Orthogonal Multiple Access   \\ 
DoS              & Denial-of-Service                             & Non-DTN          & Non-Delay Tolerant     \\ 
DTN              & Delay Tolerant                                & NTN              & Non-Terrestrial Networks   \\ 
DVB              & Digital Video Broadcast                       & O3K              & On-Off Keying       \\ 
ECEF             & Earth-Centered Earth-Fixed                    & OPEX             & Operating expense    \\ 
ECI              & Earth-Centered Inertial                       & ORAN             & Open Radio Access Network  \\ 
EDRS             & European Data Relay System                    & PFD              & Power Flux Density      \\ 
EIRP             & Equivalent Isotropically Radiated Power     & QKD              & Quantum Key Distribution \\ 
EPFD             & Effective Power Flux Density                  & RAS              & Radio Astronomy Service    \\ 
ESA              & European Space Agency                         & RF               & Radio Frequency  \\ 
FCC              & Federal Communications Commission (USA)     & RTD              & Round Trip Delay     \\ 
FDMA             & Frequency Division Multiple Access          & SDN              & Software-Defined Networking   \\ 
FSO              & Free Space Optical                            & SIC            & Successive Interference Cancellation \\ 
FSS              & Fixed Satellite Service                       & SIGINT           & Signals Intelligence   \\ 
GPS              & Global Positioning   Satellite                & SIN              & Space Information Network  \\ 
GSO              & Geostationary Orbit                           & SMN              & Space Mobile Network    \\ 
HAPS             & High Altitude Platform Station                & SNG              & Satellite News Gathering  \\ 
HEO              & Highly Elliptical Orbit                       & SNR              & Signal-to-Noise Ratio   \\ 
IAU              & International Astronomical Union              & TDMA             & Time Division Multiple   Access  \\ 
IOL              & Inter-Orbit Link                              & TDMA             & Time Division Multiple Access   \\ 
IoT              & Internet of   things                          & TESS     & Transiting Exoplanet Survey Satellite   \\ 
ISL              & Inter-Satellite Link                          & UAV              & Unmanned Aerial Vehicle    \\ 
ITU              & International Telecommunication Union         & UCS              & Union of Concerned Scientists    \\ 
JAXA             & Japan Aerospace Exploration Agency            & VSAT             & Very Small Aperture Terminal  \\ 
\hline
\end{tabularx}
\end{table*}

% Section III. NGSO  classification
%========================================
\section{NGSO System Characteristics and Classification}\label{sec:ngso_classification}
%========================================
We can differentiate two categories of NGSO satellite systems, as described by International Telecommunication Union (ITU): (i) the early systems that were designed to provide voice and low-rate data services, and (ii) the recent NGSO constellations that were introduced for provisioning global broadband services. In the first category, Iridium, Globalstar, and Orbcomm are the three projects that became operational and started service in late 1990s; despite, these systems went through bankruptcy around the year 2000, but later they have survived and are still operational \cite{Butash2021}. Typically, the frequency bands of the mobile satellite service (MSS) were used; namely, portions of L-band and S-band were assigned for uplink and downlink to enable the satellites to provide service globally \cite{Restrepo1996}. The second category of NGSO constellations is in competition with the high-throughput satellites. Specifically, multiple projects have been already announced in this category but there exists only two operational systems to date, i.e.  O3b and Starlink \cite{Huang2020a}. Further, OneWeb was one of the early projects that launched more than 70 satellites, which also survived bankruptcy in 2020. Another early project was LeoSat that planned to deliver high-speed Internet using 108 satellites, which folded in 2019 due to lack of investment \cite{Braun2021}. Additionally, these modern systems use frequency bands of the fixed satellite service for the user links, i.e. the Ku- and Ka/K-bands. There is also a possibility to add higher frequencies in the future for some systems, where even more bandwidth is available. More information about the current frontrunner projects for the communication with a constellation of satellites will be presented in the next subsection.

In addition to the aforementioned intrinsic features and advantages of NGSO satellites, there are more motives for the rising interest in NGSO constellations over the traditional GSO systems. 
Particularly, since NGSO systems require a large number of satellites to provide uninterrupted service such systems offer consequently a very high throughput and spectral efficiency \cite{Pachler2021}.
Further, the communication through the satellite constellations can bypass the terrestrial network infrastructure when they are connected via inter-satellite links (ISLs) for routing communication data in space, which will definitely improve the privacy of data transmissions \cite{Manulis2020}. In addition to the reduced signal propagation delays in NGSO communication systems comparing to GSO, low orbit constellations with ISLs have also lower delays than terrestrial fiber-optic systems since the speed of light in vacuum (free space) is approximately 50\% higher than in a fiber-optic cable ($3 \times 10^8$ versus $2 \times 10^8$ m/s) \cite{Handley2018}. Moreover, since some of the NGSO constellations (operational and planned) utilize non-equatorial orbits, they naturally can cover higher latitudes than GSO satellites \cite{Ashford2004}.

Accordingly, these advantages have increased the involvement of NGSO satellites in plentiful applications, such as telecommunications, Earth and space observation, navigation, asset tracking, meteorology, and scientific projects. In this section, we classify NGSO systems depending on the provided services into two groups: space-based Internet providers and space missions as follows.

%NGSO satellites have been already used in numerous applications, such as telecommunications, Earth and space observation, asset tracking, meteorology, and scientific projects. Depending on the provided services, NGSO systems can be classified into two categories: space-based Internet providers and space missions.

\subsection{Space-based Internet Providers}
NGSO space-based Internet providers aim to provide high-speed low-latency Internet access competitive with terrestrial broadband communications. This will not only empower satellite communications to compete for long-distance backhaul and mobile users but also address underserved populations, where currently only 39\% of the world’s population have access to terrestrial broadband infrastructure \cite{ITU2020}. Thus, NGSO space-based Internet systems can reach the developing world where it is financially unfeasible to lay fiber-optic networks. Additionally, high-latitude populations in some regions such as Alaska, northern Canada, and Russia can be served by these space-based Internet systems, which are currently served by a poor terrestrial communication infrastructure. Further, many advantages and enhancements can be achieved by employing NGSO space-based Internet systems to serve the growing broadband requirements of maritime and aeronautical services \cite{Graydon2020}.

\begin{figure}[!t]
\centering
\includegraphics[width = 0.48\textwidth]{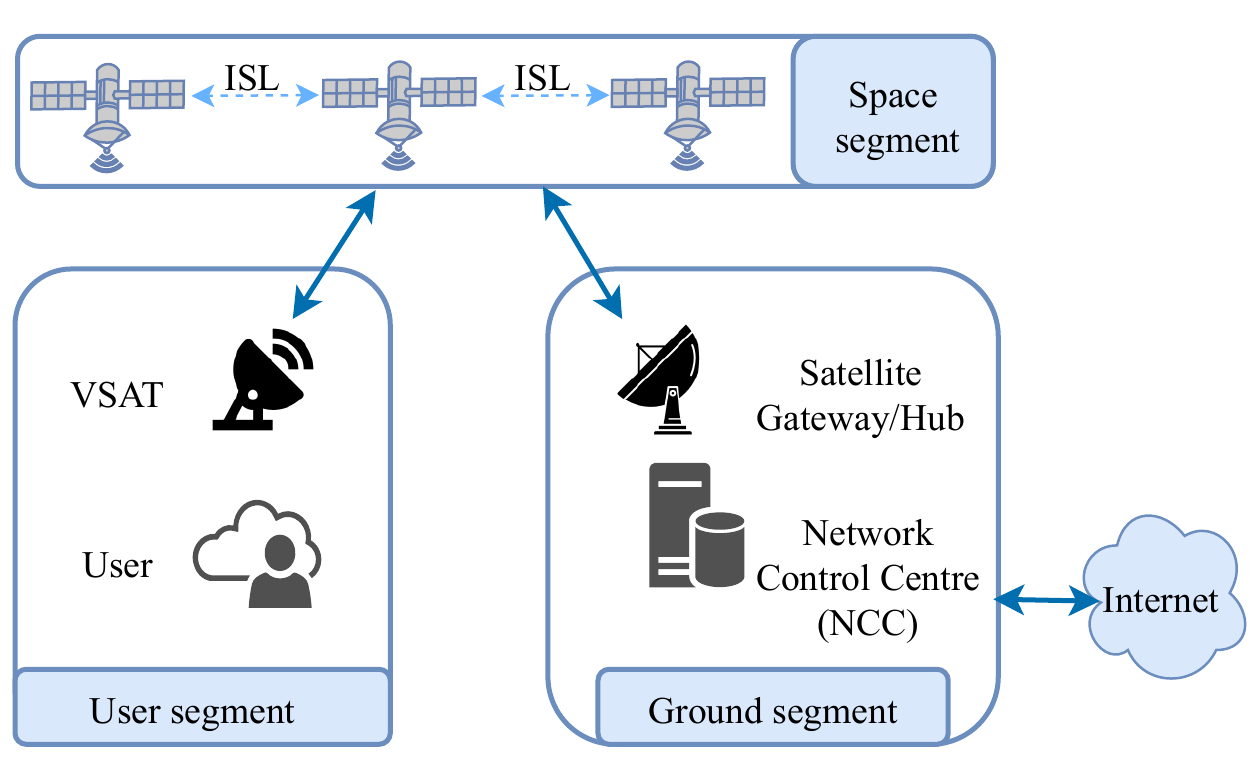}
\caption{Schematic diagram for a space-based Internet system.}
\label{fig:space_Internet}
\end{figure}

A space-based Internet system generally consists of three main components: space segment, ground segment, and user segment (see Figure \ref{fig:space_Internet}). The space segment can be a satellite or a constellation of satellites, while the ground segment involves a number of ground stations/gateways that relay Internet data to and from the space segment, and the user segment includes a small antenna at the user location, often a very small aperture terminal (VSAT) antenna with a transceiver. Additional critical entities within this structure are (i) network management centre (NMC) and (ii) network control centre (NCC) \cite{Maral2009}. The centralized NMC is the functional entity in charge of the management of all the system elements such as fault, configuration, performance, and security management. The NCC is the functional entity that provides real-time control signalling such as session/connection control, routing, access control to satellite resources, etc. \cite{DVBRCS}.

%A space-based Internet system generally consists of three main components: space segment, ground segment, and user segment (see Fig. \ref{fig:space_Internet}). The space segment can be a satellite or a constellation of satellites, while the ground segment involves a number of ground stations/gateways that relay Internet data to and from the space segment, and the user segment includes a small antenna at the user location, often a very small aperture terminal (VSAT) antenna with a transceiver. Additional critical entities within this structure are (i) network operation center (NOC) and (ii) satellite control center (SCC). The centralized NOC manages the terminal population and coordinates the information exchange with external networks, while SSC monitors and configures the network through tracking and command links.
%Other components include a satellite router which allows the subscribed user to utilize Internet services, and a centralized network operations center (NOC) which controls all communications over the satellite link. 

The space-based Internet services have been in use for several years now, but only for a limited number of users, and most of the existing systems utilize GSO satellites (e.g. SES, Inmarsat, Viasat, Eutelsat) \cite{Sachdev2009}. However, it is well known that the latency is one of the main impairments in  GSO communication systems in addition to the high propagation path loss.
This is also the reason why GSO-based Internet systems cannot be used for particular services that require a low latency connectivity, and why NGSO satellites are becoming more popular for high-speed broadband services. In addition, being closer to Earth means that signal propagation path loss is low and requires smaller antennas at the user side, which allows to serve new types of users. Among these providers we summarize some major satellite mega-constellations as follows.
\begin{itemize}
    \item Starlink of SpaceX: Starlink constellation is expected to contain nearly 12000 satellites in the initial phase with a possible later extension to 42000 \cite{Spacenews2019}. The first 12000 satellites are planned to orbit in three different altitudes above Earth: 1440 in a 550 km altitude, 2825 at 1110 km altitude and 7500 satellites at 340 km.
    Regarding the space segment, the satellites have four phased array antennas of approximately equal size to serve the Ka/K band beams, with separate antennas for reception and transmission.
    %As of today, SpaceX has launched the first 1035 Starlink satellites mainly orbiting at 550 km in planes inclined at 53$^{\circ}$ \cite{SpaceflightNow2020}. 
    Each satellite will carry a regenerative payload with a phased array antenna which will allow each of the beams to be individually steered towards the on-ground users. The minimum elevation angle for a user terminal to communicate with the satellites is 40$^{\circ}$, while the total throughput per satellite is envisioned to be 17- 23 Gbps, depending on the characteristics of the user terminals \cite{Portillo2019}. 
    %In fact, SpaceX is already delivering Internet services and has started a testing phase, termed as “better than nothing (beta)”. The delivery of beta services is quickly expanding and covering more and more regions on Earth due to the periodic launch of new satellites joining the Starlink constellation. The first feedbacks coming from users that are already utilizing the beta services report data speeds varying from 50 – 150 Mbit/s and latency from 20 – 40 ms \cite{Starlink}.
    
    \item OneWeb: Satellite constellation of OneWeb will comprise 648 satellites by the end of 2022 according to the latest updates, among which 110 are already launched \cite{OneWeb2021}. In this initial phase the satellites will be placed in 18 circular polar orbit planes at an altitude of 1200 km, where each plane is inclined at 87$^\circ$ \cite{Portillo2019}. 
    %The most recent constellation plan released by OneWeb states that the total number of satellites in the future will reach 6372 \cite{OneWeb}. This is a dramatic reduction compared to the previous plans of having a constellation of 47884 satellites in orbit. 
    OneWeb space segment will have a transparent bent-pipe payload with non-steerable, highly-elliptical user beams. Their coverage on Earth will guarantee that every on-ground user will be within the line-of-sight (LoS) of at least one satellite communicating at a minimum elevation angle of 55$^\circ$. In addition, each satellite will contain two steerable gateway antennas, where one of them will be active, while the other will assist as a back-up for handover procedures \cite{Portillo2019}.
    
    \item O3b of SES: This provider was founded in 2007 and stand for the “other 3 billion”. Its aim is to provide Tier 1 Internet connections to the developing countries, as often they are constricted by their international connections. The space segment architecture of O3b is based upon 20 satellites (started with the launch of four satellites in 2013) in equatorial circular orbit at an altitude of 8000 km delivering low latency fibre-like connectivity to any area approximately 45 degrees north and south of the equator with offering user-level broadband services at around 500 Mbit/sec \cite{O3b2021}. In this system, twelve reflector antennas are mounted on the largest spacecraft surface, two antennas are used for gateway links, while the other 10 support user links. These antennas are moving constantly to track the spots they are intended to serve on ground.
    SES has also O3b mPOWER project that initially comprises 11 MEO satellites to provide multiple terabits of global broadband connectivity for applications including cellular backhaul to remote rural locations. The O3b mPOWER satellites use steerable spot beams that can be shifted and scaled in real-time to fulfil users' needs, and they will operate in conjunction with the existing SES fleets. Fig. \ref{fig:O3b} shows an example of a multi-orbit space-based Internet provides similar to the O3b mPOWER constellation pattern. 
    \begin{figure}[!h]\centering
    	\includegraphics[width=0.45\textwidth]{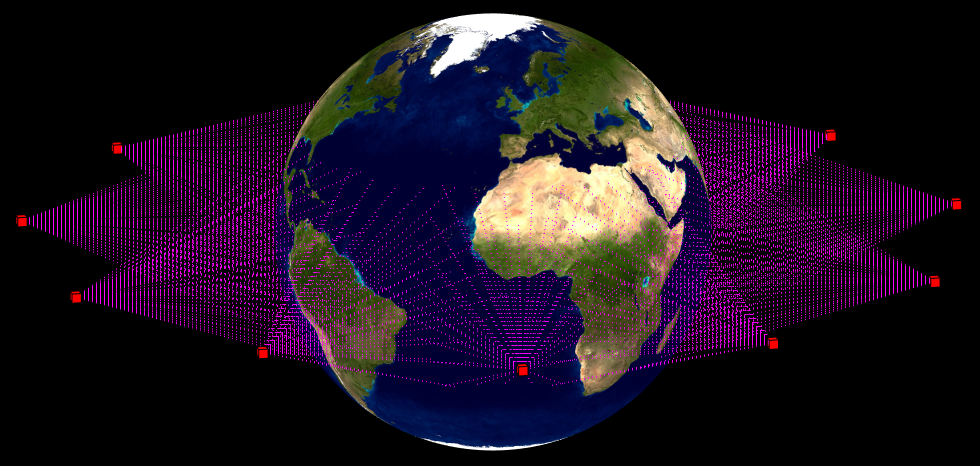}\vspace{-2mm}
    	\caption{Constellation topology of a multi-orbit NGSO system.}
    	\label{fig:O3b} \vspace{-2mm}
    \end{figure}
\end{itemize}

In addition, many private sector companies worldwide foresee market opportunities to extend their services via NGSO constellations. For instance, Amazon plans to launch over 3,000 LEO satellites through ``Project Kuiper'' to offer high-speed broadband connectivity to people globally \cite{Kuiper}. LeoSat is launching a constellation of up to 108 satellites to provide data communications in the challenging polar regions of the world. Telesat LEO plans to have 177 satellites and has already received an initial license to start providing service in Canada. Boeing also plans to have 2,956 satellites in orbit and 1,396 satellites will be launched within the first 6 years. Huawei plans to build a 10,000 satellite LEO constellation called Massive VLEO for beyond 5G systems, where a low satellite altitude of 300 km will be used for ultra-reliable low-latency communications, the large number of satellites will cover the massive machine-type communications and broadband communications \cite{Huang2020a}. These are not all the involved companies in this rapidly growing market and listing all of them is beyond the scope of this paper.

\subsection{NGSO Space Missions}
Space has become more affordable and accessible than ever due to the recent evolution of satellite technologies and the emergence of small satellites; namely, in addition to the traditional players in space sector, any country, university, startup or even school can now reach space in an affordable way and within short periods. Thus, the sky is not the limit any longer, where these developments have unlocked the missions that satellite can carry and execute for different needs and applications. In this subsection, we outline and concisely present some of the current and most relevant space missions in the context of NGSO satellites.
 
 \subsubsection{Earth and Space Observation}
 
One of the most widespread uses of satellite constellations in different orbits is capturing high-resolution images of Earth and outer space as the current technology makes it possible to have latest-generation cameras that fit perfectly to the size of small satellites \cite{ruf2018new}. On one hand, NGSO satellites have made far-reaching enhancements in the field of cartography to provide accurate and up-to-date maps, from the most remote to the most populated areas on Earth \cite{corbane2021}. On the other hand, utilizing small satellites to obtain information and images of outer space is attracting more attention to search for transiting exoplanets and space exploration. For instance, NASA has launched transiting exoplanet survey satellite (TESS) system in 2018 in its missions for searching for planets outside of our solar system \cite{Mackereth2021}, including those that could support life. Another proposals proceeding in this field involve the use of small satellites as guide star for latest generation telescopes, which require steady references to explore and capture quality images of exoplanets and celestial bodies \cite{allen2020}.

\subsubsection{Asset Tracking}

One of the main NGSO satellite fields is asset tracking owing to their capability of ensuring a stable and precise service with a complete coverage anywhere on the planet. Satellite payload in asset tracking projects consists of a device equipped with communication components to collect information sent from objects on ground and to transmit it back to ground stations \cite{Davarian2020}. The main practical applications of NGSO missions in this field include but not limited to: 
\begin{itemize}
    \item Fleet management where satellite tracking of all types of vehicles such as cars, trucks, buses, industrial machinery, etc. Lower orbit constellations also have the ability to strengthen wireless networks and provide solutions for precise control of vehicles and mobile resources, even in inaccessible areas \cite{padilla2020m2m}.
    
    \item Logistics companies can track their enormous amount of goods and products in real time and can estimate the time of arrival of any product regardless of its price by using NGSO satellite constellations. Small satellite constellations are an effective solution to improve security, control and traceability in the logistics sector by tracking containers, goods, and machinery that may require controlled transport conditions (e.g., temperature and movement) and different means of transport to reach their destinations (e.g., road, rail, airplane, ship) \cite{hao2019design}.
    
    \item Maritime tracking to ensure the safety of each type of vessels and to control some problems that often affect maritime traffic can be improved with help of NGSO small satellites. Additionally, in areas of low coverage with limited access of terrestrial networks, small satellites can be helpful to ensure at all times the location and control of vessels \cite{te_hayder2020}.
    
    \item Aircraft tracking to obtain accurate information in seconds in different areas is already existed using terrestrial systems. However, most of the recent and biggest air tragedies regarding to disappearance of planes have happened in shaded areas. To avoid such issues, NGSO small satellite-based solutions for automatic dependent surveillance – broadcast (ADS-B) systems can be very helpful to increase safety, improve air traffic control, receive certain information provided by flight sensors in real time and know at all times the exact location of the aircraft \cite{te_hayder2020}.
 \end{itemize}

In this perspective, the private venture Spire Global \cite{Spire} operates a large multi-purpose constellation of nanosatellites for tracking the maritime, aviation and weather patterns. They collect and offer datasets include Automatic Identification System (AIS) data that contains the movements of ships and vessels across the world, ADS-B data that constructed from tracking airplanes across global airways, and real-time weather conditions. Additionally, instead of observing the Earth in the visible domain using cameras, KLEOS \cite{kleos} as private company is utilizing LEO satellites to locate radio transmissions from different devices, a sort of reverse GPS. This radio-frequency-mapping can benefit the maritime market for locating ships that may have lost connection with their transponders. They also offer locating dark, unseen, obscured, obfuscated, covert maritime activity that may indicate activities such as illegal fishing and trafficking. 

\subsubsection{Scientific and Environmental Missions}

Missions in this category involve very broad applications and experiments in space within a wide range of disciplines, and the objective of each mission determines the payload of satellites \cite{Moreira2015,Gutiérrez2020}. NGSO satellite can facilitate some missions that employ small satellites as summarized in the following use cases:
\begin{itemize}
    \item Meteorology: NGSO Small satellites can play a significant role in storm detection and in the development of climate and weather models that enhance weather forecasts. For instance, RainCube project (Radar in a CubeSat) of NASA has already entered the testing phase for the location, tracking and analysis of rain and snowstorms all over the planet \cite{Davoli2019}.
    \item Agriculture: Crop monitoring is another potential use of small satellites, where a better control of harvests, the improvement of the quality of agricultural products, the finding of diseases in crops, and analysis of the ramifications derived from the periods of drought can be accomplished by using NGSO satellites \cite{Onojeghuo2018,ping2018mini}. 
    \item Educational activities: The development of scientific experiments outside the Earth has become another common application of small satellites, which are unprecedented opportunities brought up by NGSO small satellites with their countless possibilities \cite{buscher2019}.
    \item Environmental protection: Several projects can be conducted in this context based on small satellite, such as detection and monitoring of forest fires, studying the progress of melting ice, fighting against ocean pollution, detection of oil spills and spills, monitoring of marine life, controlling of desertification, along with other initiatives \cite{Jarrold2020}.
 \end{itemize}

\subsubsection{Government Space Programs}

Small satellite developments have backed the so-called space democratization after some many years of controlling the space by a handful of countries, as it is now reachable by not only companies and startups, but also countries that want to launch their space programs or to expand their current capabilities \cite{Skinner2020}. The goals of these government programs varies from national security to emergency response. For example, small satellite can be used for signals intelligence (SIGINT) \cite{weinbaum2017sigint} by monitoring the radio electric and electromagnetic spectrum, identifying signals from the Earth and space, observing communication traffic patterns, detecting interference and locating its origin, preventing the illegal use of radio bands and unauthorized emissions. Moreover, in crisis and natural disasters such as Earthquakes, tsunamis or hurricanes, small satellites can help to act quickly, to immediately know the degree of the damage and to manage relief and rescue teams. Additionally, some existing applications of small satellites in tackling potential threats from outer space have focused on the study and possible diversion of potentially dangerous asteroids for our planet, such as the Hera project of the European Space Agency (ESA) \cite{Hera}. In space exploration missions, small satellites are gradually gaining prominence, e.g. NASA's InSight mission has already sent nanosatellites to travel into deep space to provide real-time telemetry of the spacecraft landing on Mars \cite{zweifel2021seismic}.

Beyond the aforementioned features and applications, NewSpace will continue to be an endless source of new research and application opportunities. Besides, many promising technical advances are anticipated to emerge in the future satellite systems that will boost NGSO constellations and small satellites for more practical applications. For instance, an important advance is the introduction of artificial intelligence (AI) to space networks \cite{Fourati2021}. In addition to enabling automatic learning systems using AI for satellite constellation management, intelligent ground station networks will optimize the control and operation of such a massive and diverse system architecture.

%=====================================================================
%  Section V communication prospects 
%=====================================================================
\section{NGSO Communication Prospects} \label{sec:ngso_communication}
Basically, 
%communication is the procedure of establishing a connection between two points for information exchange, and
a communication system serves to transfer information through a channel extends from the transmitter to the receiver \cite{Schwartz1996}.  
Due to the typical limitations of the terrestrial wireless communications in terms of coverage and capacity, it appears extremely challenging, if not impossible, to provide a global wireless connectivity with sufficient quality of service 
%The traditional ground wireless communications have limitations in terms of capacity and coverage, which means depending only on the terrestrial communications systems will not assure a full global wireless connectivity, 
especially in harsh environments such as ocean and mountains \cite{Hubenko2006}. Alternatively, satellites have the ability to serve distant locations by redirecting the signals received from a transmitting device on Earth back via a transponder, i.e. satellites can establish a communication channel between a transmitter and a receiver at different locations on Earth. Thus, satellite systems have the capacity to extend communication coverage to isolated or remote islands and communities, and fulfilling the needs of areas and countries with limited infrastructure investments \cite{Minoli2009}. Therefore, it is critical to utilize miscellaneous communication systems and architectures to accommodate the increasing growth in the number of users and services in various scenarios and applications \cite{Zeng2016}.

At the moment, satellite communication systems are going through a profound change due to the rise of NGSO constellations alongside with the existing GSO satellites \cite{Deng2021}. Specifically, GSO systems are in constant contact with ground stations where these stations control the GSO operations, while NGSO systems will need to be built on more autonomous and reconfigurable architectures, and the assumption of persistent contact with ground stations is no longer feasible in the NGSO setup \cite{Wang2021}.
This impediment inflicts several critical issues upon the communication framework of NGSO satellite constellations. 
%The communication aspects of this evolving field have not been entirely covered in the existing works. 
Thus, this section focuses on the key research progress for utilizing NGSO satellites to further advance the communication systems. Through this, we will start discussing physical layer technologies and radio access schemes, and then, moving forward to explore the networking aspects. Next, the overall system characteristics and architectures of the evolving NGSO constellations will be presented. 
%followed by providing some futuristic views for NGSO implementation scenarios and interesting applications.  

\subsection{Physical and Radio Access Aspects}
Physical connectivity and accessing the multi-orbit satellites are crucial factors that seriously affect the communication quality and system performance. Therefore, we focus in this subsection on understanding the relevant physical/link procedures and features including antenna system  and link budget analysis along with reviewing  the recent developments concerning inter-satellite connectivity,  waveform design, and link diversity and multiplexing.

%---------------------------------------------
\subsubsection{\textbf{Antenna Systems}}\label{antenna}
%---------------------------------------------
The multi-beam antenna and the phased array are mainly used in NGSO satellites, which provide a large number of higher-gain small beams, and thus, increase system capacity over the entire coverage area.
Additionally, frequency and polarization are commonly reused within the generated beam patterns.  The direct-radiating array is also employed on NGSO satellites owing to its wide scanning angles and the better off-boresight performance (lower scan loss) than that provided by a phased array antenna \cite{Braun2021}. In this context, the effective isotropic radiated power (EIRP) is a parameter characterizing transmit antennas, which is equal to the product of the transmitted peak power and antenna maximum gain, and hence, it represents one of the driving parameters for the design of a satellite link budget \cite{Maral2020}. A procedure to estimate the NGSO satellite EIRP is given by the ITU in the recommendation ITU-R S.1512 \cite{ITUR_2001}. Furthermore, the choice of antenna type and specifications for a particular application have many drivers such as orbit, carrier frequency, beam size, flexibility, edge-of-coverage gain roll-off, and cost.

Additionally, active antennas have been also evaluated to be used for lower orbit satellites, which are antennas encompass active electronic components like transistors in opposite to the passive antennas that simply consist of inactive components such as metal rods, capacitors and inductors \cite{Natera2012}. Thereby, introducing active antenna system technology to NGSO satellites opens up new opportunities for developing flexible multi-beam payloads and for exploiting massive MIMO techniques in satellite communications \cite{Chen2020}. 
With the active antenna system technology, it is possible to have more controllable antenna ports as baseband, which will offer higher spatial degree-of-freedom for deploying flexible  beamforming technology \cite{Pavan2015}. These advances empower satellite platforms to serve moving user terminals and enable more swift multi-orbit interoperability. 

In the past, pointing to multiple satellites or to different orbits was limited to governmental and institutional users, security and defense satellite applications and business dedicated networks mainly due to the high cost and size of user antenna \cite{Lueschow2020}. However, the new low-cost, electronic-array-based flat-antenna are expected to be a game changer for expanding the role that NGSO satellites play in connecting devices, with little installation, configuration and maintenance effort \cite{Ovejero2021}. With its rapid switching speeds and agility to track and switch seamlessly and reliably between satellites and constellations, flat-antenna arrays enable the exploitation of the essential advantages provided by the combination of multiple constellations by proper beam steering and interference nulling capabilities.    
    
Similar to the terrestrial systems, multiuser precoding and detection techniques are expected to be widely adapted in NGSO systems, where they could be used either by regenerative onboard processors or ground end-to-end from bent-pipe satellites \cite{Sharma2020a}. 
Both user terminals and satellites can use active phased-arrays antenna for transmission and receiving to overcome the propagation loss. 
In this context, massive MIMO can substantially increase degrees of freedom, enhance spectral efficiency, and achieve high data rates \cite{massiveMIMO}. However, massive MIMO allows distant beams to reuse frequency, which may bring about inter-beam interference due to the non-zero side lobes. Therefore, side lobe suppression technologies are required for the use of massive beamforming in NGSO satellites \cite{Chen2020}. Generally, the available studies to investigate MIMO technology in NGSO satellite systems are limited. In \cite{You2020}, the deployment of massive MIMO in LEO satellites is studied with considering the LEO satellites are equipped with uniform planar arrays of antennas to serve ground users through precoding and user grouping based on statistical CSI. Authors of \cite{Feng2020} have modeled ground gateway stations and visible LEO satellites as a bipartite graph and proposed a maximum matching based solution to select the satellites that could be connected to every ground station considering basic MIMO concepts to deal with this multi-connectivity. In \cite{Goto2018}, the capacity of LEO-MIMO systems is analyzed considering the Doppler shift and allocating different channels for data and control signals. 
        
To further advance this interesting integration, more studies to exploit and explore other aspects and capabilities will be beneficial such as investigating the network architecture, channel estimation, precoding, inter-user interference, etc. Additionally, inter-satellite communications may consider massive MIMO within the high frequency bands to realize high-speed data transmission and flexible network architecture but that requires accurate and fast channel estimation. Angle-of-arrival (AoA) is promising technology in this setup for improving the multiplexing gain and radio link quality especially in the regions with  poor signal strength \cite{almamori2020}. Moreover, the high density of mega-constellation along with the simultaneous visibility of more NGSO satellites can be exploited to establish cell-free massive MIMO architecture to benefit from the efficient duplexing technique, pilot assignment, and handover management, and then, a substantial performance improvement can be achieved \cite{Abdelsadek2021}.

%---------------------------------------------
\subsubsection{\textbf{Link Budget}}
%---------------------------------------------    
Miniaturizing satellites in NGSO systems imposes restrictions on the payload design, and hence, limiting the transmit power and antenna aperture, which directly affects satellite link budget. To quantify the link budget differences between GSO and NGSO satellites, the recent 3GPP technical notes in \cite{3GPP38821v16,3GPP38811v15} are used and the results are presented in Table \ref{tab:link_budget}. Specifically, two types of user terminals and frequency bands are considered, i.e. an handheld in the S-band and a VSAT in Ka-band. For other scenarios and configurations, the interested reader can refer to \cite{3GPP38821v16,Guidotti2020a,Thales2019}. Clearly, one can observe that the obtained very low carrier-to-noise ratio (CNR) at the handheld terminal is very poor when GSO satellite is employed. In contrast, LEO link achieves a better CNR at the handheld terminal. For the VSAT terminal,  both GSO and LEO links have good link budgets owing to the high antenna gain and transmit power of the VSAT terminal but still LEO link outperforms the GSO link by up to 16 dB in uplink.
    
NGSO communication links have lower signal losses and smaller propagation delays comparing to the GSO links thanks to the lower orbits \cite{Pahl1994}. In fact, these advantages can be exploited in several ways such as miniaturizing the user equipment, reducing user terminal consumption power, increasing the spectral efficiency, and targeting latency-critical applications \cite{Lopez_2020}. Furthermore, this may allow a smooth adoption of commercial off-the-shelf modems as user terminals (e.g. smartphones and terrestrial IoT devices) to seamlessly work with NGSO satellites \cite{Rodiger2021}. In order to quantify these advantages, we calculated the round trip delay (RTD) for both GSO and NGSO communication systems. The results are depicted in Table \ref{tab:link_budget}, where an LEO satellite at an altitude of 600 km is considered as an example for the NGSO systems. Obviously, the RTD in the LEO link is about 36 times lower than in the GSO link, which is a big difference.

The mobility of NGSO satellite brings about a variable receive power at the ground terminals, which is then represented as a function of the ground antenna elevation and the slant path through the atmosphere \cite{Kallfass2021}. Besides, the high mobility of NGSO satellites  causes the well-known Doppler phenomenon and its potential impact on communication links. Doppler effect makes a time-varying frequency offset and that will complicate the channel estimation process and increase the need for high channel estimation overheads. Several approaches have been proposed in the literature to overcome this issue. For example, a state-space method proposed in \cite{Pedrosa2021} for tracking channel variations for satellite links with high Doppler frequency shifts. Reference \cite{Khan2020} has developed an analytical framework for statistical characterization of Doppler shift in an NTN where LEO satellites provide communication to ground users. Likewise, another challenge resulting from satellite mobility is the time-varying visibility of NGSO satellites, which can be relaxed by different techniques such as a proper constellation planning, \cite{Babich2020}, and design a visibility matrix with a time-varying satellite topology\cite{Lee2021}.
    
\begin{table}[t]
\caption{NGSO and GSO system parameters for numerical link budget evaluation.}
	\label{tab:link_budget}
	\footnotesize
	\centering
% 	{\setlength{\tabcolsep}{0.2em}
% \setlength{\extrarowheight}{0.1em}
		%\begin{tabular}{|l|p{0.29 \textwidth}|}
		\begin{tabular}{|p{0.16 \textwidth}|c|c|c|c|}
		\hline
				\textbf{Parameters} & \multicolumn{4}{c|}{\textbf{Values}} \\
				\hline
				\textbf{User Terminal} & \multicolumn{2}{c|}{\textbf{VSAT}} & \multicolumn{2}{c|}{\textbf{handheld}} \\
				
				 & \multicolumn{2}{c|}{\textbf{ (Ka-band)}} & \multicolumn{2}{c|}{\textbf{ (S-band)}} \\
		\hline
		        Satellite orbit      & GSO & LEO & GSO & LEO \\
		      
		\hline
		        Elevation angle (degree)  & 30 & 30 & 30 & 30\\
		 \hline
		 RTD in forward link (ms) & 515.18 & 14.35 & 515.18 & 14.35\\
		 
		 \hline
		 \multicolumn{5}{|l|}{\textbf{Downlink}} \\
		 \hline
		 Frequency (GHz)               & 20 & 20 & 2 & 2 \\
		 \hline
		 Bandwidth (MHz)               & 100 & 100 & 10 & 10 \\
		 \hline
		  Free space loss (dB) &	210.20	& 179.10 &	190.20	& 159.10\\
		 \hline
		 Atmospheric loss (dB)	& 0.52 &	0.52 & 0.07 &	0.07\\
		 \hline
		Shadowing margin (dB)	& 0.00 &	0.00 & 3.00 &	3.00\\
		 \hline
		Scintillation loss (dB)	& 0.30	& 0.30 & 2.20	& 2.20\\
		 \hline
		 EIRP-satellite (dBm)  & 90 & 54 & 99 & 74\\
		 \hline
		 G/T-user (dB/K)  & 15.86 & 15.86 & -31.62 & -31.62\\
		 \hline
		 \textbf{CNR (dB)}   & \textbf{13.44} & \textbf{8.54} & \textbf{0.51} & \textbf{6.61}\\
		 \hline
		 \multicolumn{5}{|l|}{\textbf{Uplink}} \\
		 \hline
		 Frequency (GHz)               & 30 & 30 & 2 & 2 \\
		 \hline
		 Bandwidth (MHz)               & 100 & 100 & 1 & 1 \\
		 \hline
		  Free space loss (dB) &	213.73	& 182.62 &	190.20	& 159.10\\
		 \hline
		 Atmospheric loss (dB)	& 0.50 &	0.50 & 0.07 &	0.07\\
		 \hline
		Shadowing margin (dB)	& 0.00 &	0.00 & 3.00 &	3.00\\
		 \hline
		Scintillation loss (dB)	& 0.30	& 0.30 & 2.20	& 2.20\\
		 \hline
		 EIRP-user (dBm) & 76.21 & 76.21 & 23.01 & 23.01\\
		 \hline
		 G/T-satellite (dB/K) & 28 & 13 & 19 & 1.1\\

		 \hline
		 \textbf{CNR (dB)}   & \textbf{8.28} & \textbf{24.39} & \textbf{-14.86} & \textbf{-1.66}\\
	   	\hline			
		\end{tabular}
% 		}
\end{table}

%---------------------------------------------
\subsubsection{\textbf{Inter-satellite Links}}
%---------------------------------------------
ISLs play an important role in the formation of satellite networks especially for NGSO systems. They enable command, control, communication and information processing with real time or near real time communication capabilities as well as to reduce the network dependency on the ground stations \cite{Chen2021}. Efficient ISLs will allow future space missions to be autonomous space systems.
Radio frequency (RF) and optical links are the two primary communication media for an ISL. RF has the advantage of mature technology, and does not require a tight acquisition and tracking functionalities but it suffers from interference and provide low data rate compared to optical media. The concept of using RF ISLs has been around for about 30 years \cite{keller1996link,Stuart1996}. Mororola's Iridium system is the first commercial satellite system to use RF ISLs, showing that they are practical on a large network of LEO satellites. Besides, many other Earth observation missions have used the RF ISLs as a communication way between different satellites \cite{ESAPRISMA, ESAProba3}.

In this context, Terahertz (THz) band communications are anticipated to support a wide variety of ISLs \cite{Akyildiz2020}, such as the satellite cluster networks and inter-satellite backbone networks \cite{Suen2015}. Unlike ground THz communications that suffer from short distance transmission limitations due to the atmosphere attenuation, deploying THz communications in space applications in the atmosphere-free environment circumvents this constraint and achieves high-speed long-distance links between satellites. However, there are still a number of open challenges for THz satellite communications particularity in terms of semiconductor technologies. For example, it is prohibitively difficult to produce high power THz transmitters and current THz receivers prone to higher noise figures. Thereby, with more research efforts dedicated for developments of high power THz transmitters, highly sensitive receivers, and adaptive antenna arrays, many THz communication opportunities can be explored within the NGSO satellite deployments \cite{Yu2021}.

On the other hand, free space optical (FSO) communication links have the advantage of higher data rates, smaller size, and lower power, but needs more complex acquisition and tracking functionalities \cite{Gong2020}. Two additional advantages can be added for laser-based FSO which are the low probability of intercept and intrinsic high-gain due to narrow-beam nature of laser beams. For satellite communication, FSO links have already been experimented by the ESA and  Japan Aerospace Exploration Agency (JAXA) for satellite-to-satellite link within the SILEX research program (Semiconductor Inter-Satellite Laser Experiment) \cite{tolker2002orbit}. In \cite{garcia2002preliminary, toyoshima2007results}, ground stations have been developed for optical space-to-ground links to investigate data transmission through the atmosphere. Whereas, an optical link between an aircraft and a GSO satellite was established and used to demonstrate a communication link in strongly turbulent and dynamic environment in \cite{cazaubiel2006lola}. Reference \cite{smutny20095} has considered introducing coherent modulation techniques to achieve higher data rate links  connecting LEO satellites. 

FSO technology is currently gaining momentum not only in experiments and demonstrations but also for commercial purposes in the context of connecting space missions. For instance, the European data relay system (EDRS) project utilizes optical inter-satellite link for data relay systems over multi-orbit satellites  \cite{6934556}. To react to this reality, the consultative committee for space data systems (CCSDS) has defined recently new specifications to deal with coding and synchronization of high photon efficiency links \cite{CCSDS1-19}. CCSDS is also targeting the coding and synchronization layer of a waveform supporting optical LEO direct-to-Earth links and which will rely on optical on-off keying (O3K) providing channel data rates from few Mbps up to 10 Gbps \cite{CCSDS2-20}.

From what precede, it is obvious that the evolution of FSO technology is very similar to the fiber optics a decade earlier where the latter was based on single-mode transmission and direct detection \cite{neumann2013single}.
 More interestingly, introducing quantum cryptography,  or quantum key distribution (QKD), to satellite systems for offering highly secure applications is also giving momentum to FSO links \cite{Vu2020}.
Accordingly, for future space-based FSO research topics, it will be interesting to investigate recent technologies adopted in state-of-the-art fiber optics as coherent modulation formats, multiplexing schemes, coherent receiver techniques and advanced digital signal processing at receiver and transmitter, especially, for ground-to-space and space-to-ground links where the propagation environment is challenging mainly due to the presence of the atmosphere. In short, the adoption of such advanced techniques can pave the way to new types of architectures and services, which probably will lead to new satellite communication paradigms.

%---------------------------------------------
\subsubsection{\textbf{Waveform Design}}
%---------------------------------------------
Waveform design is a critical and fundamental aspect in defining the wireless communication standards \cite{Sanctis2015}. Current satellite communications have been standardized according to Digital Video Broadcasting-Satellite (DVB-S) for both physical and link layers in GSO and NGSO systems. Specifically, the second generation DVB-S2 and its extensions DVB-S2X \cite{DVBS2X} are widely implemented due to its ability to adapt to changing propagation conditions. DVB-S2(X) includes a high number of modulation and coding schemes from which the system can select the most suitable one based on the link Signal-to-Noise Ratio (SNR). Optimal waveform design for improved transmission efficiency has been investigated within DVB compliant scenarios for GSO systems \cite{8746876}. In particular, the problem of inter-beam interference management has received significant attention, particularly for techniques implemented at the transmitter side (i.e. gateway). The reader is referred to \cite{7811843} for a detailed discussion of precoding schemes supported by DVB-S2(X).
    
However, NGSO systems have emerged with a focus on  particular promising 5G satellite use cases and associated requirements such as latency-sensitive applications \cite{Gaudenzi1327}. The key driving factors proposed by the research community to meet the heterogeneous requirements of 5G-NR are new candidate waveforms for flexibly rendering the waveform parameters. Nevertheless, the typical satellite channel impairments, such as variable propagation delay, high Doppler shift, high non-linear degradation, impose designing challenges on the physical layer to support NR operations. 
Satellite communications community is currently investigating alternatives to facilitate the integration of NTN into the 5G ecosystem. In this direction, ESA is currently running a research project for 5G enabled ground segment technologies over the air\cite{ESA_5GGoa}, which is devoted to investigate the necessary modifications in the 5G-NR standard to enable the direct radio access of terrestrial communication networks via satellite. Indeed, direct access from legacy user terminals is constrained by the low-power wide-area network technologies. Similarly, the joint project of 5G Space Communications Lab \cite{5GSpaceLab} aims at implementing a space communications and control emulation platform for the next-generation of space applications including the evaluation of different small satellite formation control and cooperation configurations for NTN-5G networks.
    
Additionally, new air interface waveforms and numerologies are being analyzed in \cite {Jayaprakash2020} within the ongoing activities and studies of 3GPP related to the feasibility and standardization of necessary adaptations for the 5G NR to support integrated-satellite-terrestrial networks with LEO satellites. Reference \cite{Oltjon2021} studies and analyzes the random access procedures over NTN-based 5G systems and the challenges imposed by the increased signal propagation delay.
The adaptability of candidate waveforms under satellite channel impairments is evaluated in \cite{Jayaprakash2019 }. The impact of the satellite channel characteristics on the physical and medium access control layers in terms of transmitted waveforms is assessed in \cite{Guidotti2019}, particularly random access, timing advance, and hybrid automatic repeat request in the context of satellite-based NR networks.
In \cite{Arapoglou_20202}, the feasibility of direct broadband access from NGSO systems to low gain handheld user equipment is studied in the millimeter wave (mmWave) range, from a regulatory, user equipment characteristics, space segment, link budget and system point of view. The aforementioned works have identified major challenges in terms of waveform design that require more research efforts to realize NGSO integration with 5G-NR standards.

%---------------------------------------------
\subsubsection{\textbf{Access Design and Multiplexing}}
%---------------------------------------------
One of the most important enablers of vigorous NGSO satellite communications is the efficiency of radio access schemes. Many access solutions for heterogeneous terminals with stationary and non-stationary channel characteristics have been developed in the framework of terrestrial networks \cite{Bai2019}. Herein, radio access design for on ground and airborne users is more complicated and challenging compared to the terrestrial case due to the different relative motion of those users with respect to satellite nodes, variable propagation delay, uneven transmit powers, link availability, and variable QoS profiles \cite{boero2018}. Accordingly, to simultaneously serve a large number of heterogeneous users and provide ubiquitous and flexible connectivity solutions, it is imperative to devise efficient techniques that provide fair radio access and scheduling in order to avoid collisions, interference, and imbalanced capacity distribution \cite{Mirza2019}.

Radio access techniques can be generally classified into (i) random access (RA) schemes (also known as uncoordinated multiple access protocols) and (ii) coordinated access schemes \cite{clazzer2017}. The first class allows a set of users to transmit over a common wireless medium opportunistically and independently \cite{Wong2018}. In contrast, the second class requires  a central unit (typically a ground station) to coordinate users accessibility and to allocate a dedicated non-interfering resource to each transmitter. 
On one hand, coordinated multiple access schemes are mainly suitable for communications that require high data throughput, high levels of QoS, and relaxed users power consumption. 
On the other hand, the uncoordinated solutions can be generally employed when user terminals have tight power consumption restrictions, limited time of network visibility, and relatively low data rate and QoS requirements \cite{wu2020massive}.\\

%Further, coordinated access schemes can be efficiently employed in  static communication networks, i.e. the number of users and their throughput requirements are constant \cite{clazzer2017}. On the other hand, the uncoordinated solutions can be mainly employed in dynamic communication environments to serve a very large number of users that have limited transmission cycles \cite{wu2020massive}.\\

\noindent     
\textbf{Random Access}\\
The RA dates back to the 1970s when the ALOHA protocol was developed to solve the problem of interconnecting university computers located in different Hawaiian islands \cite{abramson1970}. Since then, several developments have been proposed to employ the ALOHA protocol in different scenarios. A review of RA techniques with particular emphasis on the challenges and the possible solutions applicable to satellite networks is provided in \cite{de2018random}. The slotted Aloha scheme is studied in \cite{s21217099} as a medium access control technique for multiple users under the coverage of a constellation of LEO satellites. In this setting, the throughput and packet loss rate are analyzed while considering potentially different erasure probabilities at each of the visible satellites within the constellation.
Further, a novel framework of analysis of diversity framed slotted ALOHA (DFSA) scheme with interference cancellation for RA satellite platforms is proposed in \cite{Pietro2020}. Designing a reliable RA preamble and detection scheme for high-dynamic LEO scenarios is considered in \cite{Zhen2021} to effectively enhance the radio access efficiency. A framework of non-orthogonal slotted Aloha (NOSA) protocol is proposed and analyzed in \cite{Qiwei2018} for achieving high user throughput IoT-oriented satellite networks.

In addition to the slotted Aloha, unslotted solutions have been utilized for satellite based IoT applications, e.g. an unslotted spread spectrum Aloha scheme is applied to LEO based IoT transmissions in \cite{Pansoo2019}. To support long-range connectivity and scalability, LoRa (Long Range) and Sigfox methods are proposed as unslotted Aloha protocols. Specifically, LoRa is based on spread spectrum techniques \cite{LORA2013} and Sigfox is used for ultra-narrow band transmissions \cite{Asad2021}. For example, the work in \cite{zhang2022} has proposed a low complexity orthogonal LoRa algorithm for multiple users occupying the same frequency bandwidth in order to improve the multiple access performance of satellite IoT services. Further advances have been recently  introduced to the unslotted Aloha protocols such as LoRa MAC adaptability for LEO satellites \cite{rs13194014} and enhanced spread-spectrum ALOHA with successive interference cancellation (SIC) \cite{mengali2018modeling}.
%, and adapted to the satellite environment \cite{herrero2012high, 7880912}  
%Additional works \cite{rs13194014, ferrer2019review, palattella2018enabling} have studied the LoRa MAC adaptability for LEO satellites. Other advanced unslotted Aloha protocols are also proposed recently as an alternative to the classical ones, e.g. enhanced spread-spectrum ALOHA with SIC (E-SSA) \cite{mengali2018modeling}, and adapted to the satellite environment \cite{herrero2012high, 7880912}.}

%Moreover, LoRa Cloud has been recently developed by Semtech corporation as an addition to the LoRa protocol to be applied to edge device-to-cloud geolocation platforms for global asset tracking\cite{lora_LEO}.

\noindent 
\textbf{Coordinated Access}\\
In the coordinated multiple access, the system assigns the available resources to the accessible users in a dedicated systematic way \cite{rom2012}. This class includes some typical paradigms that can be breifly descried  as follows:
\begin{itemize}
\item Frequency division multiple access (FDMA) divides the channel bandwidth into multiple orthogonal frequency subchannels, and assigns each subchannel to a certain user.
\item Time division multiple access (TDMA) utilizes non-overlapping time slots to serve different users.
\item Code division multiple access (CDMA) serves multiple users simultaneously over the same time slot and frequency band but with different codes.
\item Space-division multiple access (SDMA) employs spatial beamforming techniques to serve multiple users at the same time and using the same frequency band. Inter-user interference can be mitigated using the directional beamforming.
\item Non-orthogonal multiple access (NOMA) grants access to multiple users simultaneously via non-orthogonal resources. The fundamental principle behind NOMA technology is the classical multi-user information theory.
\end{itemize}

These coordinated multiple access protocols have become more mature and applied into various satellite scenarios.
Specifically, several works on coordinated multiplexing protocols have considered the large number of satellites to be deployed as a satellite sensor network, and then applied the concept of terrestrial wireless mesh networks to satellite nodes and space missions. Moreover, the work in \cite{Radhakrishnan2016} has conducted a survey on the classical multiple access protocols highlighting their benefits and pitfalls from efficiency and scalability perspectives. Authors in \cite{Radhakrishnan2016} have also proposed two access schemes for a distributed network of small satellites; namely, (1) a modified carrier sense multiple access (CSMA) scheme that establishes communication only when it is required, and (2) hybrid time TDMA/CDMA protocol where multiple satellites from different clusters utilize same time slot using different codes. Furthermore, an efficient transmission scheme for flexible multiple access to several small LEO satellites has been proposed in \cite{Katona2020} based on a QoS-aware scheduler. Further, applying the SDMA scheme in mobile satellite communications is investigated in \cite{Ilcev2011} taking into consideration the transmitting power and bandwidth to achieve efficient simultaneous communications among multiple mobile satellite users.

Furthermore, NOMA scheme can be incorporated in the NGSO multi-beam satellite architecture to design efficient transmission strategies that aim at increasing radio access flexibility and capacity \cite{Lin2019}. Specifically, NOMA is more suitable for multi-layer multi-orbit NGSO systems owing the inherited near-far Effect resulted form to the relative motion of satellites, the spatial distributions of users, and the various received power levels. Employing NOMA requires performing SIC at the receiver side, and hence, NOMA will alleviate co-channel interference, accommodate more user terminals, and improve system spectral efficiency \cite{islam2016power}. 
NOMA was investigated within multi-beam satellite systems as a radio access approach in \cite{perez2019non} to maximize system capacity while  taking into account the precoding methods, power allocation, and user grouping schemes. In \cite{Zhixiang2022}, NOMA-based average age of information (AoI) in LEO satellite-terrestrial integrated networks is considered to improve average AoI performance with considering transmission delay, queuing delay, and propagation delay.

All the above-mentioned access methods have limitations when serving a massive number of users. In particular, the RA schemes may cause too many collisions, and similarly, the overhead required for the coordinated multiple access to manage an enormous number of users will overwhelm the system. To address this issue, massive access is an emerging technology can be utilized in such cases to accommodate the number of users per transmission medium by a high order of magnitude compared to current methods \cite{Jiao2021}.
In this domain, a massive MIMO transmission scheme with full frequency reuse (FFR) for LEO satellite communication systems is proposed in \cite{You2020}. Utilizing MIMO beamforming for massive connectivity in NGSO systems offers a seamless accessibility but requires employing multi-user detection approaches at the receiver, such as  SIC and joint processing of signal copies received by multiple NGSO satellites \cite{Zhao2021}. Applying NOMA and massive MIMO to LEO satellite communication systems in studied in \cite{Zhixiang2021} to improve spectral efficiency.
Thereby, granting access for a massive number of diversified users to the NGSO satellites while taking into account the relative motion among different entities, variable QoS requirements, differential delays, and Doppler effects are an interesting research direction to be further developed in order to satisfy the growing demand for NGSO satellite services.

In this context and motivating by the fact that a user terminal can see multiple NGSO satellites at the same time, there is an opportunity to combine the signals from multiple satellites for improving the aggregated data rate, beam load balancing, and improving the robustness of the satellite link exploiting path diversity. In this direction,  the concept of carrier aggregation (CA) can be also adopted \cite{CA_Mirza,Hayder2020} to the NGSO communication systems. CA is a well-developed technique in Long Term Evolution-Advance (LTE-A) standard for terrestrial networks and has succeed to significantly boost the performance through maximizing the spectrum utilization and satisfying the extremely high throughput requirements in certain circumstances \cite{3GPP2012}. Besides, in the context of NTN networks to extend the coverage of uplink transmissions performed by users towards NGSO satellites, the supplementary uplink technique \cite{Feng2020a} can be utilized here to enhance user experience. Supplementary uplink and dual connectivity can also be used for reducing latency or providing higher uplink data rates in power-limited situations \cite{Dazhi2022}.

In brief, a continuous and ubiquitous wireless coverage can be attained via NGSO satellites, which can be seen as a cost-effective solution for reliable coverage across different geographies. However, NGSO network scalability requires more research and development in the technologies related to physical and radio link layers in order to reap the benefits of the improved connectivity offered by the low orbit satellites. In particular, dynamic beamforming through active antenna arrays can improve link performance, while FSO communication and THz bands are envisioned for ISLs to achieve high data rates on the order of Terabits per second. Besides, the design of waveform is also critical in this context as it impacts the transmission efficiency. Further, multiple access schemes such as NOMA, MIMO, and CA can be incorporated in the NGSO multi-beam satellite architecture to grant radio access to the foreseeable large number of users.

% Multi-orbit architecture maybe!!
%==================================
\subsection{Networking Aspects}
\begin{figure}[!t]\centering
	\includegraphics[width=0.5\textwidth]{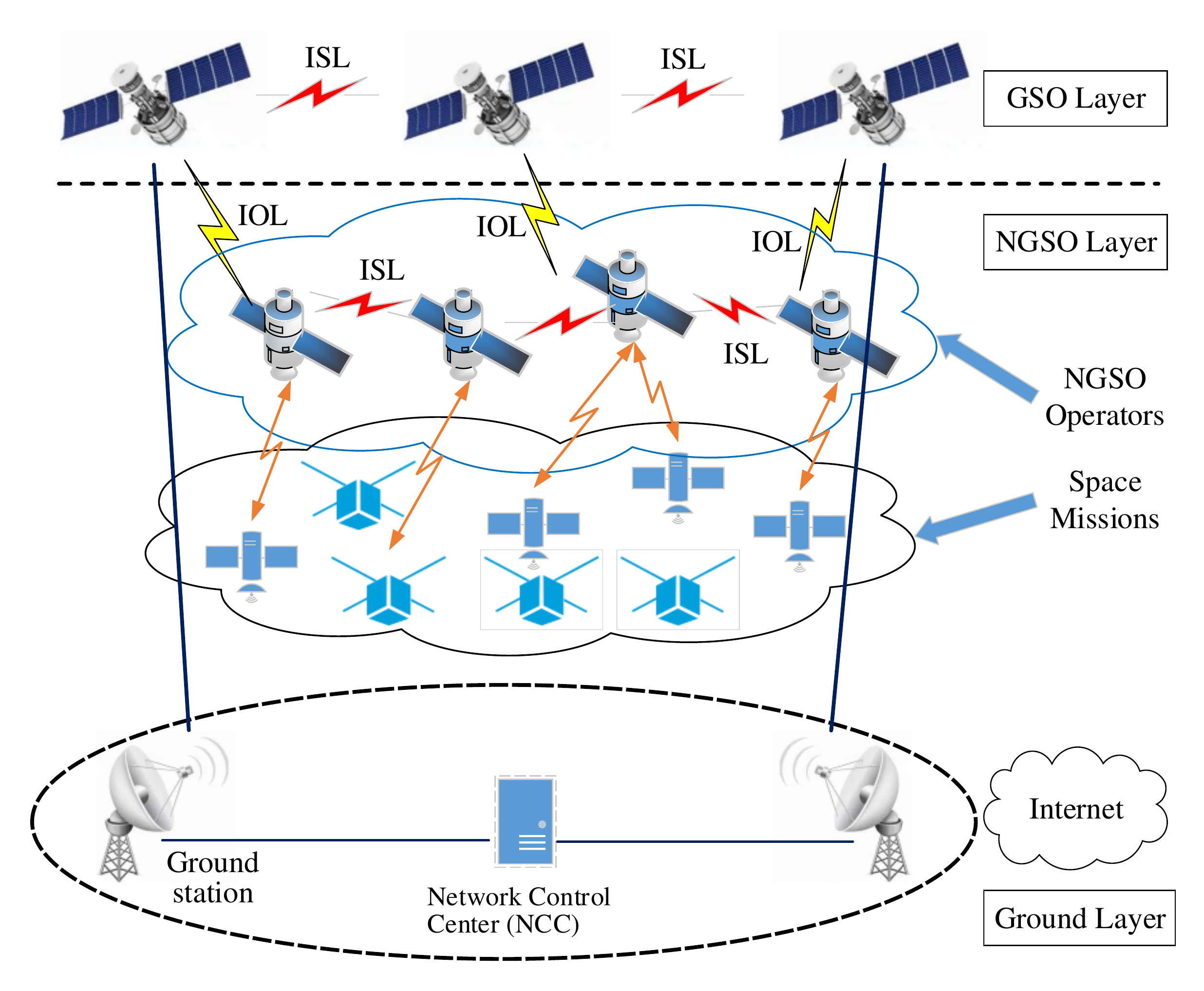}%\vspace{-4mm}
	\caption{General schematic diagram of a multi-layer space information network.}
	\label{fig:SIN_diagram}
\end{figure}

%==================================
The challenge inflected by the ever-growing NGSO systems and mega-constellations that are launched for various space applications is the necessity for a real-time uninterrupted connectivity, which is fairly infeasible in current satellite system infrastructure due to the magnitude and cost of the needed gateway network on ground. Thereby, there is a demand for the development of new space network infrastructures to supplement and extend the satellite communication systems \cite{Wei2016,Dai2012}. To this end, satellites can be deployed as a space information network (SIN) using ISLs and inter-orbit links (IOLs) as illustrated in Fig. \ref{fig:SIN_diagram}. Establishing SIN architectures is more economically efficient and more suitable for the heterogeneous integrated satellite communications. SINs can fulfill the increasing complexity of application requirements, and can also eliminate the use of the excessive number of gateways. This architecture is particularly favorable for the areas where acquiring gateway sites is difficult \cite{Hassan2020}. However, this expansion leads to numerous theoretical and technical challenges considering the restricted spectrum, energy, and orbits resources. Therefore, the fundamental issues for nurturing the development of SINs need to be explored and adequately addressed.   
%---------------------------------------------
\subsubsection{\textbf{Space Information Networks}}
%---------------------------------------------
SINs are integrated networks based on different space platforms including GSO and NGSO satellites, and airships on high altitude platform stations (HAPS) to provision real-time communications, massive data transmission and processing, and systematized information services \cite{Sheng2019}. 
Furthermore, SINs enable communication and cooperation between satellites for traffic routing, throughput maximization, latency minimization, and seamless coverage \cite{Guo2021}. Similarly, SINs can provide coordination and awareness of the operational characteristics about each counterpart system, and thus, achieve a successful coexistence between different satellites without imposing detrimental interference to their concurrent transmissions \cite{Yu2016}. However, the expected better performance of space-based networks will be achieved at the cost of higher complexity that is essential for load balancing between satellite links and for finding paths with the shortest end-to-end propagation delay.

Unlike terrestrial networks, SINs consist of various, independent, and complex components that are designed for different purposes. The high complexity and variety of satellites along with their diverse portfolio of constellations and the high-speed mobility of NGSO with respect to the Earth's surface impose exceptional technical challenges on the system design and communication environment. 
To this end, EDRS project of ESA is dedicated to the development and implementation of data relay satellites that are placed in GSO orbit to relay information to and from NGSO satellites, spacecraft, and fixed ground stations that otherwise are not able to permanently transmit/receive data \cite{EDRS}. Similarly, NASA has also invested in this concept by developing the so-called space mobile network (SMN) to be an analogous architectural framework for near Earth space applications \cite{Israel2016}. In parallel, some works in the literature consider connecting lower orbit satellites with other higher orbit ones for routing data packets and reducing the dependency on the ground stations. For instance, the concept of system of systems was introduced in \cite{Walker2010} to study the availability and capacity of a simplified scenario consists of a few multi-orbit satellites. In \cite{Guo2020}, an architecture has been proposed based on fog environment via considering the underutilized moving satellites as mobile fog nodes to provide computing, storage and communication services for users in satellite coverage areas.

%---------------------------------------------
\subsubsection{\textbf{In-space Backhauling}}
%---------------------------------------------
The deployment of SIN requires developing more sophisticated traffic distribution schemes to manage the growing number of satellite nodes and users to achieve network congestion control, resource utility maximization, energy efficiency, and resilience structures \cite{Jia2021}.  Interestingly, the aforementioned satellite advancements allow on-board regeneration and Layer 3 routing that render satellites to active network elements rather than simple bent-pipe relays \cite{Jian2019}.  In particular, in-space backhauling is a crucial part in this setup along with designing  
efficient routing mechanisms that consider the unique features of the multi-layered multi-orbit SINs. 
In this configuration, several challenges imposed at the satellite network level related to dense satellite distribution, transmission delays, QoS priorities, uneven distribution of data flows, and the dynamic change of the network's topological structure. 
Designing efficient in-space backhauling protocols starts from evaluating the infrastructure parameters such as topology variation, bandwidth, link delay, in addition to traffic generation profiles of the heterogeneous user services/classes and computational and storage capabilities of the nodes.

Furthermore, utilizing NGSO systems can be extended beyond the rural and remote areas to include the urban areas where satellites can provide an alternative backhaul solution.  In 5G systems, the backhaul demands inherent in networks with large numbers of small cells can be accommodated via NGSO satellite networks to be used as a single centralized backhaul for traffic offloading, edge processing, and resource sharing \cite{Turk2019}. In fact, satellite-based backhaul communication can be seen on the horizon within the standardization efforts in 3GPP associated with identifying the technical requirements and solutions to support NR-NTNs \cite{3GPP38821v16}. In this context, backhaul connection solutions of terrestrial 3GPP-based infrastructure have been investigated in \cite{ Liolis2019} to enable ubiquitous 5G coverage with integration of satellite infrastructure of the existing satellite network operators. In this setup, performance of a terrestrial-satellite system can be improved by considering dynamically varying backhaul capacity determined by the satellite selection and backhaul capacity optimization \cite{Di2019}.

%---------------------------------------------
\subsubsection{\textbf{Software-defined Networking}}
%---------------------------------------------
In the context of satellite communications, researchers have already developed several routing algorithms under the satellite network constraints. Traditional routing schemes have been used in distributed and centralized systems depending on the network topology and mission requirements. These approaches require each satellite to store the entire network topology along with the routing tables \cite{Radhakrishnan2016}, but that is difficult in complex SINs and consumes more power and bandwidth. In parallel, it has been extensively concluded that an effective solution is given by the well-known paradigm software-defined networking (SDN) \cite{xu2018software}. SDN paradigm enables dynamic, programmatically efficient network configuration in order to improve network performance, management, and monitoring. Therefore, SDN has a tremendous potential to succeed in SINs owing to its capability to implement a reactive scheme for end-to-end traffic engineering development across both terrestrial and satellite segments. 

In the literature, prior works in \cite{yang2016seamless} and \cite{sheng2017toward} have proposed to distribute an SDN controller on the ground, while some other works have considered the placement of the SDN controller on GSO satellites \cite{li2017service}. As intermediate solution, an SDN-based infrastructure for multi-layered space terrestrial integrated networks is introduced in \cite{shi2019cross} to distribute the SDN controller entities among GSO satellites, terrestrial infrastructure, and HAPS, which is still seen as a terrestrial-dependent SDN network. Furthermore, it has been emphasised in \cite{akyildiz2019Internet} that there is a lack of SDN-based architecture solution specifically designed for small satellites, where all the prior works mainly focus on the traditional LEO, MEO and GSO satellites. Authors in \cite{akyildiz2019Internet} have presented a detailed SDN structure adapted to the Internet of space things and small satellites but their implementation is more applicable for monitoring and Internet provisioning for remote areas, which make the developed platform terrestrial-dependent. Thereby, an efficient SDN-based architecture for multi-layer SINs requires more developments to achieve a flexible framework capable of facing the dynamicity of the nodes and the heterogeneity of the traffic.

%---------------------------------------------
\subsubsection{\textbf{Network Slicing}}
%---------------------------------------------
NGSO networks are expected to grow largely in size and complexity due to the wide adoption of services and users. In addition, the combination of terrestrial and satellite networks has introduced new dimensions of network heterogeneity and dynamicity. Hence, network management is a critical challenge to provide NGSO satellite communication services in a more flexible, agile, and cost effective manner. Therefore, embracing network slicing concept through adopting network virtualization and softwarization technologies can significantly increase the degrees of freedom in the network management process \cite{Drif2021}. Network slicing is envisioned as a promising design approach within the multi-layer NGSO network structure owing to its ability of enabling optimal support for wide-reaching heterogeneous services that share the same radio access network. Network slicing is made possible thanks to SDN and network function virtualization (NFV) technologies \cite{ahmed2018demand}. With SDN, networks can be dynamically programmed through centralized control points, while NFV enables cost-efficient deployment and runtime of network functions (e.g. computing, storage) as software only \cite{Rost2017}. Through this paradigm, satellite networks can be seamlessly integrated with other heterogeneous networks in a 5G ecosystem.

Network slicing enables running multiple logical networks as independent tasks on a common physical infrastructure, where each network slice represents an independent virtualized end-to-end network and allows operators to perform multiple functions based on different architectures \cite{Ferrus2016}. However, applying the network slicing paradigm to NGSO satellites is not a straightforward task and provokes a number of challenges. For instance, assigning dedicated spectrum resources to individual slices can diminish the multiplexing gains due the scarcity of radio spectrum \cite{Wu2021}. Besides, satellite service providers will need to carefully plan and apply different technologies to serve diverse users with considering radio access heterogeneity and spatial diversity. Network slicing works efficiently when more information can be provided by the infrastructure about the shared parts to the network slice but exposing such information creates new potential security vulnerabilities between infrastructure providers and their partners. In other words, network slicing is still at an early stage of its application into 5G systems and requires novel algorithms and solutions to involve the NGSO systems.

Concisely, the development of new space network infrastructures to complement and expand the satellite communication systems has several advantages for both terrestrial networks and NTNs, which will improve communication and cooperation between different network nodes for traffic routing, throughput maximization, latency minimization, etc. However, some prospects in this evolving part are still rather overlooked and require more research efforts for integrating NGSO satellites with the global communication infrastructure. Specifically, multi-orbit SINs can improve organization and operation of all network entities while providing in-space backhauling for diverse services and applications. Likewise, SND is another promising technology that has the features of flexibility, programmability, and effective control, which can improve network resource utilization, simplify network management, reduce operating cost, and boost the evolution and innovation. Moreover, the combination of network slicing technology and SINs can offer autonomous network resource allocation and flexible management mechanisms.

%============================================
\subsection{System and Architectural Aspects}
\begin{figure}[!t]
	\centering
	\includegraphics[width=0.5\textwidth]{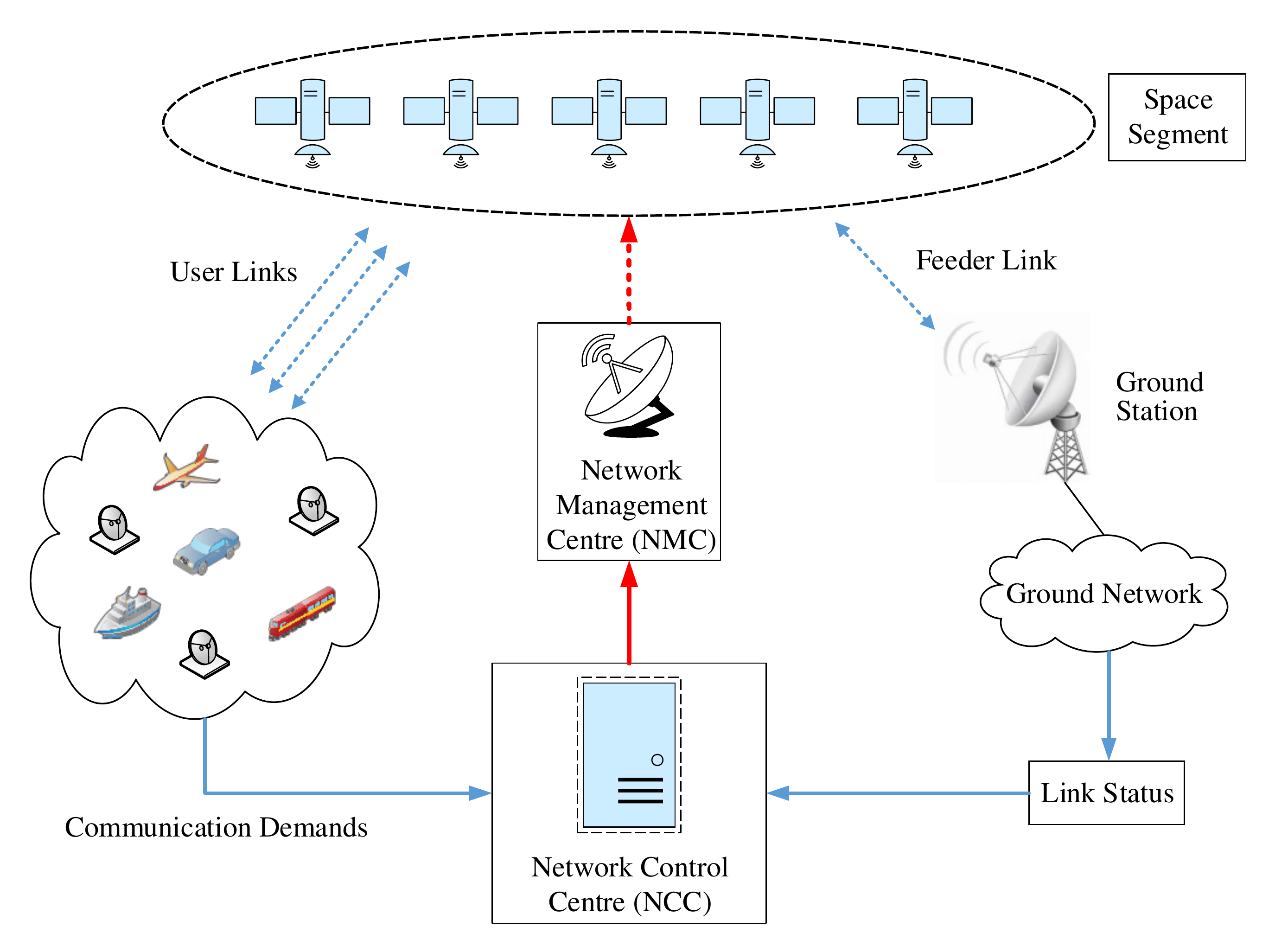}
	\caption{Diagram of a satellite communication system architecture.}
	\label{fig:sat_architecture}
\end{figure}

%============================================
Satellite systems are very complex cyber-physical systems, which are challenging to operate due to the immense physical distance with the asset. Traditionally, GSO satellites can be operated individually, since each asset occupies a specific orbital slot and provides service over a specific coverage area. The operation is usually split between two main functions NMC and NCC \cite{DVBRCS}, as presented in Fig. \ref{fig:sat_architecture}. The two types of operations are tightly linked and there are strict coordination procedures between the two, especially when the communication payload has to be reconfigured (e.g., carrier switching, power control, etc.). Furthermore, the relevant hardware and software for NMC and NCC are usually replicated over multiple geographically distanced sites on the globe to avoid single points of failure on the ground.

For NGSO systems, it is apparent that these operations become even more involved for two main reasons: a) a large number of gateways is required, b) there are multiple satellites that have to be jointly operated/configured so that they optimize the performance of the communication service as the constellation rotates. The former reason is currently a large capital expenditures (CAPEX) for the deployment of mega-constellations, which can be partially mitigated by deploying ISLs for routing communication data in space \cite{Bacco2019a}. The latter reason is mainly driven by the relative motion between the constellation and user terminals, and unbalance of data traffic/demand depending on the geographical location of the users, which requires the constant reconfiguration of satellites in terms of resource allocation \cite{joroughi20195g}.

The control and operation mechanisms are fundamental issues for the NGSO satellites. These issues can be settled by operating NGSO system in either centralized or decentralized manners \cite{Palattella2021}. In centralized architectures \cite{zheng2019}, high efficient network management can be achieved but that comes at the expense of incurring a non-negligible complexity and an increased operating expenditures (OPEX). Specifically, network controllers in the centralized architectures typically execute in servers located at a terrestrial network. The control channels between a controller and each node (satellite or ground station) will require additional bandwidth resources in addition to the resource allocation burden. On the other hand, in decentralized architectures, each NGSO satellite independently regulates its operating parameters such as power allocation and topology management \cite{russell2002}. This architecture requires the development of energy-efficient and delay-sensitive distributed algorithms that are able to run in the on-board units of satellites such that the amount of messages that need to be exchanged among satellites and their neighbors is limited.
However, global optimal control and operation policies are difficult to achieve in this decentralized manner.

Additionally, the high heterogeneity and complexity of NGSO systems alongside with the high-speed mobility with respect to the Earth's surface inflict multiple system and architectural challenges that need to be carefully addressed \cite{Hayder2021}. Particularly, NGSO systems have to confront the interference issues due to the coexistence with other satellite systems and terrestrial networks, which requires developing efficient interference coordination/mitigation techniques. Likewise, the new features of NGSO satellites with their heterogeneous resources are exacerbating the resource management challenges. Thus, resources management strategies that are cognizant of the topographies of different satellite systems are indispensable in such dynamic propagation environments. In addition, the integration of NGSO satellite systems into Internet infrastructures comes with serious security threats due to the large constellations that will include hundreds or even thousands of satellites providing direct connectivity. Thus, the essential system requirements to achieve smooth and reliable NGSO communications are discussed in this subsection including resource optimization, interference management, spectrum sharing, and security issues. 
%---------------------------------------------
\subsubsection{\textbf{Resources Management and Optimization}}
%---------------------------------------------
In order to satisfy the growing traffic demand a thorough design of the resource allocation strategies with respect to power, bandwidth, time interval, beam and antenna (to exploit spatial diversity), needs to be done, cf. \cite{lagunas2021}. Nevertheless, the demand satisfaction is much more challenging with NGSO compared to GSO satellites because of less available resources due to a much smaller payload \cite{Vidal2020,Lagunas2020}. Also, the complexity requirements of the employed algorithms are much more strict with NGSO satellites, since the optimization parameters quickly become outdated. These requirements may even pose a burden for the feasibility of optimization, since the resource management problems are often non-convex and have many optimization parameters, which require iterative convexification methods to obtain reasonably good solutions. Thus, it might be useful to reduce the number of parameters or apply low-complexity metaheuristics and machine learning methods \cite{9237970}.

Resource management is significantly affected by the employed satellite coverage scheme \cite{kuang2017}. One of the two popular coverage schemes can be adopted by NGSO systems: (i) spot beam coverage and (ii) hybrid wide-spot beam coverage \cite{Lutz2012}. In a spot beam coverage scheme, each satellite provides multiple spot beams to offer coverage over its service area, where their footprint on Earth's surface moves along with the satellite trajectory. This scheme is simple but the handover between beams are more frequent because the coverage area of a single spot beam is rather small. On the other hand, in hybrid wide-spot beam scheme each satellite provides a wide beam for the whole service area and several steering beams for users employing digital beamforming techniques. The spot beams are always steered to the users, and thus, the provided footprint is nearly fixed during the movement of satellite. In this scheme, handover occurs only between the wide beams of adjacent satellites, such that the number of handover operations substantially decreases due to much larger beamwidth. Another approach suggested in \cite{trunking2019} involves joint optimization of the available resources, number of beams and beam width. Through this, it is possible to take into account the desired handover frequency as well as demand satisfaction per beam.

The overlapping coverage of multiple NGSO satellites, especially if they belong to different orbital planes, poses a challenge for the resource allocation, as explained before, since asynchronous satellites can attempt to satisfy the same demand, thus heavily interfering with each other and wasting the resources \cite{Su2019}. To avoid such scenarios, multiple adjacent NGSO satellite may need to be jointly optimized, which dramatically increases the complexity of the optimization.
Besides, the spectrum allocated to the applications served by NGSO systems is neither constant nor fully dedicated during the service interval. Specifically, the spectrum resource blocks are allocated based on the available spectrum resources, the speed requirement, and the priority of the service and user. The traditional frequency reuse schemes may not be feasible in some scenarios due to the fast handover between the adjacent beams or even adjacent satellites \cite{Wu2019}.

%---------------------------------------------
\subsubsection{\textbf{Interference Management}}
\begin{figure}[!t]
    \centering
    \def\svgwidth{210pt}
	\fontsize{8}{4}\selectfont
	\scalebox{1}{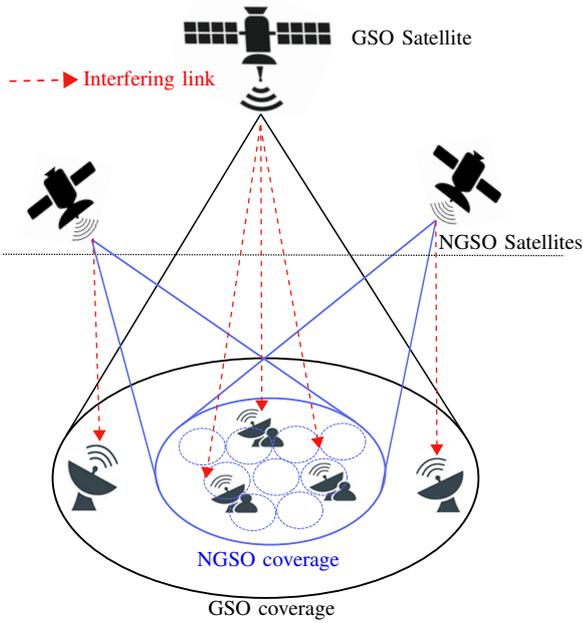}\vspace{-2mm}
	\caption{Aggregated interference scenario involving a GSO and multiple NGSO satellites.} \label{fig:NGSO_Interf}\vspace{-2mm}
 \end{figure}

%---------------------------------------------
Interference analysis of the emerging NGSO constellations should take into consideration the effect of the aggregated interference due to utilizing a large number of multi-beam satellites and applying frequency reuse techniques \cite{Tianjia2020}. For example, Fig. \ref{fig:NGSO_Interf} shows an interference scenario where multiple satellites having multi-beam and multi-carrier per beam.
Despite the several prior works on developing interference mitigation techniques for satellite systems, the high heterogeneity and ambiguity about the parameters of the emerging deployments make the effectiveness of these traditional mitigation techniques questionable when applied to NGSO. Moreover, most of the prior works focus mainly on the inter-system interference between GSO and NGSO, while the serious issue of NGSO-NGSO interference was recently  addressed only in \cite{Braun2019,Wang2018,Pourmoghadas2016,Tonkin2018}. The downlink interference between LEO system and GSO systems is analysed in \cite{Braun2019} to evaluate the inter-satellite spectrum coexistence performance. The conclusions of \cite{Braun2019} indicate that the existing spectrum regulations may be insufficient to ensure GSO protection from NGSO interference. Furthermore, \cite{Braun2019} evaluates three simple interference mitigation strategies: (i) look-aside or pointing to another LEO satellite within the visible area, (ii) band-splitting (which goes against the maximization of the spectral efficiency), and (iii) exclusion zone or avoid operation in areas where the interference is intolerable.

The authors of \cite{Wang2018} have  analyzed the coexistence of GSO and LEO constellations in Ka band with focusing on the exclusion angle strategy (i.e. LEO is not allowed to transmit in this angle) to assess the reduction in the in-line downlink interference from LEO to GSO systems. Similarly, a power control mechanism and a methodology for inter-site distance determination are proposed in\cite{Pourmoghadas2016} to minimize the interference in Ka band caused by an NGSO satellite towards a GSO system.
The impact of NGSO-NGSO co-channel interference on the achievable throughput for NGSO constellations is studied in \cite{Tonkin2018}. Band splitting interference mitigation techniques are also investigated in \cite{Tonkin2018} with considering the Ka and V bands. Accordingly, the highly heterogeneous NGSO constellation properties with the interference interactions need to be thoroughly analyzed for satellite deployments over different bands and constellations. 
    
Most of abovementioned works analyze uplink and downlink scenarios where coverage areas of NGSO and GSO satellites overlap. However, the interference between ISLs needs more investigation, which is a serious problem in the NGSO networks as it may occur not only in the overlap of coverage areas but also wherever inter-satellite communications take place. This interference scenario is more challenging and complex to analysis because of the relative motions and constellation dynamics. In \cite{Mendoza2017}, the impact of the interference generated by inter-satellite links of a LEO constellation is studied over both GSO satellites and ground stations that are part of the GSO satellite network. An interference suppression scheme is proposed in \cite{Wang2020b} based on a code-aided technique in the global navigation satellite system ISLs. Alternatively, optical ISLs have the advantage in this regards over the RF ISLs due to their robustness against interference and signal jamming.

%---------------------------------------------    
\subsubsection{\textbf{Spectrum Sharing}}
%---------------------------------------------    
The concept of mega-constellation brings about spectrum sharing challenges between NGSO and GSO systems. These mega-constellation satellites will operate at the same frequencies that are currently used by GSO satellites including the Ka and Ku bands, which has raised some serious concerns among GSO satellite operators. Therefore, coordination and awareness of the operational characteristics about each counterpart system is essential in order to achieve a successful spectrum sharing between different satellites. Spectrum sharing concept has received a tremendous
research attention to combat the spectrum scarcity issue in wireless communication networks \cite{Haykin2005}. 
Basically, a typical spectrum sharing scheme consists of a primary system with the privilege to use its licensed spectrum and a secondary system that has a lower priority and may utilize the spectrum but without causing any detrimental interference to the primary transmissions. Thereby, NGSO systems may employ this concept and exploit the spectrum allocated to GSO satellites or terrestrial networks by using underlay, overlay and interweave spectrum sharing techniques.

In the interweave scheme, NGSO systems operate in a sensing-transmitting fashion, i.e. secondary users first sense the licensed spectrum and when it is not occupied the secondary users utilize this spectrum for data transmission \cite{Goldsmith2009}. Whereas, in the underlay model, the transmit power of secondary NGSO systems is strictly constrained to satisfy the interference threshold of the primary GSO systems \cite{Zou2010,Amarasuriya2016a}. On the contrary, in the overlay scheme, the secondary NGSO system assist the primary transmissions through cooperative relaying techniques in exchange for spectrum access without posing stringent transmit power restrictions \cite{Han2009}. Integration of these spectrum sharing paradigms into NGSO communications can provide significant benefits in terms of spectral efficiency and transmission reliability \cite{Sharma2017}.

Some recent works have considered multiple spectrum sharing scenarios to wisely share spectrum resources within the coexistence of the multi-beam GSO-NGSO systems. For instance, a database-based operation is foreseen a possible approach can achieve sort of coordination between mixed satellite systems \cite{Hoyhtya2017}. Additionally, a flexible spectrum sharing approach is proposed in \cite{Gu2021} for a scenario where multiple LEO ground users are located within the coverage of a GSO satellite. In this model, the GSO satellite is considered as the primary system and the LEO satellites are the secondary system. This approach aims at optimising the throughput of LEO satellites under the premise that the QoS of GSO satellite is guaranteed. Further, a spectrum-sharing framework is designed in \cite{Wang2020c} where LEO system can work  concurrently with GSO systems in the interference region by accessing the shared spectrum in both overlay and underlay modes. In \cite{Tang2021}, the  flexibility of LEO beam hopping satellites is utilized in a spectrum sharing scenario where an LEO satellite constellation system is considered as a secondary system to share the spectrum resources of a GSO satellite.

%---------------------------------------------    
\subsubsection{\textbf{Secure Communications}}
%---------------------------------------------
Satellite communications typically rely on ground stations for securing the transmissions, which pushed the majority of security research efforts to focus mainly on the data links between satellites and the ground stations, i.e., uplink and downlink \cite{He2019}. However, the steadily growing deployment of the space-based networks shows that there will be also a big security risk in the data communication between satellites and even the internal structure of satellites. These security issues cannot be ignored and they deserve more attention.  Additionally, the complex structure of the space-based wireless network requires various security modeling and analysis for the space-based NGSO networks in combination with certain application scenarios.
    
Proper security mechanisms are essential for NGSO communication systems because they are susceptible to security threats such as eavesdropping, jamming, and spoofing. For instance, any sufficiently well-equipped adversary can send spurious commands to the satellite and gain full access to satellites as well as data, enabling them to cause serious damage. In addition to the blind jamming \cite{7397710}, intelligent jamming exploiting the communication protocols can be used \cite{jamming1}. In this context, applications of satellite-aided massive uncoordinated access are very vulnerable to such intelligent jamming due to the reduced coordination, i.e. increased uncertainty related to the structure of the received signal. Another example for potential malicious activity that requires additional security measures is related to denial-of-service attacks, which can be conducted by adversaries via sending a large number of spurious messages to the satellite \cite{Chowdhury2005}. Thus, satellites under this attack will spend significant computational  processing power and time to the spurious messages, which degrades the quality of service for the legitimate users. NGSO satellites can be particularly susceptible to this kind of attacks due to rather limited computational power, such that the satellite can be easily overloaded with processing tasks and may not be able to provide the requested service within the short visibility window.

Security of satellite communication is traditionally provisioned through cryptography-based techniques on the upper layers. The drawback of these techniques is a high computational complexity \cite{Xiao2019}. Thus,  more efficient and sophisticated methods from the areas of quantum key distribution (QKD), block-chain technology (BCT), and  physical layer security have been proposed \cite{zheng2011physical,bonato2009feasibility,secure2017,Xu2019}. 
QKD provides means to detect, if the transmission has been eavesdropped or modified. For this, the quantum coherence or entanglement is employed, which is based on a unique connection between the transmitter and the receiver. The drawback of this scheme is, however, the need to exchange the keys, which may need time, since entangled particles need to be produced and sent. Hence, this approach may not always be suitable for NGSO and especially LEO satellites due to the fast passage of the satellite.

The communications between ground stations and NGSO satellite constellations require decentralized tracking and monitoring of active and inactive space assets. In addition, it requires assessing the space environment through a network of multi and heterogeneous of satellite nodes in different orbits. In this respective, BCT can be utilized for securing satellites communications and authenticating space transactions between the NGSO constellations and ground stations \cite{Xu2019}. The key feature of BCT is to authenticate satellite's identity, ground station's identity, or communication pattern validity through a history record of changes such as configuration and re-configuration history of the satellite and space information network. Therefore, BCT can be beneficial to protect satellite communication against the denial-of-service (DoS) and distributed denial-of-service (DDoS) and insider attacks. Although, BCT challenges should be scrutinized as well, such as the BCT database storage and distribution for all satellite nodes in a network.

On the other hand, physical layer security is known to be an effective approach to achieve reasonable levels of security without imposing additional computational complexity for data encryption/decryption \cite{Hayder2017}. This technique is very popular in the terrestrial domain, where the spatial filters are designed with respect not only to the user demands, but also to the secrecy against an eavesdropper with partially known or unknown location. However, the satellite-terrestrial communication link usually does not have enough spatial diversity to distinguish between the intended users and eavesdroppers. Hence, this method applied to satellite communications is still in its infancy, cf. \cite{8850067}. Interestingly, the joint precoding over multiple NGSO satellites with overlapping coverage areas may solve this issue in some cases, since the spatial diversity associated with the antennas of the adjacent satellites can be exploited to increase the secrecy.

%Section IV Challenges 
%%%%%%%%%%%%%%%%%%%%%%%%%%%%%%%%%%%%%
\section{NGSO Deployment Challenges} \label{sec:ngso_challenges}
%%%%%%%%%%%%%%%%%%%%%%%%%%%%%%%%%%%%%
Notwithstanding the growing interest in NGSO satellites due to their essential feature of providing high-speed pervasive connectivity for a wide variety of use cases and applications, there are still many daunting challenges in the NGSO satellite evolution to be addressed in order to achieve high quality communications \cite{Hayder2021}. In this context, the 3GPP has pointed out the main challenges related to the mobility and orbital height of the satellite in Release 16 \cite{3GPP38821v16}. Afterwards, Release 17 establishes basic mechanisms to manage the identified challenges in Release 16 and provides a first set of specifications to support NTNs in complementing the 5G system along with the terrestrial networks. Release 17 builds on previous releases (15 \cite{3GPP38811v15} and 16) with the aim of improving 5G system performance, where NTN channel models and necessary adaptations to support NTN were recognized. The key difference among these potential solutions is related to which functionalities are implemented on-board satellites. More specifically, satellites can act either as relay nodes between 5G user equipment (5G-UE), or as 5G access points (5G-gNodeB) to extend 5G radio access network (5G-RAN) coverage, or as backbone/backhaul supports. Besides, the additional study in \cite{3GPP_sataccess} investigates the possible employment of satellite networks as active nodes in the 5G access operations. However, NTN integration is also leading to challenges associated with the deploying and adapting the satellite networks to the technologies that are originally designed for terrestrial networks \cite{Bacco2019}.

\begin{figure}[!t]\vspace{-3mm}
	\centering
	\includegraphics[width=0.4\textwidth]{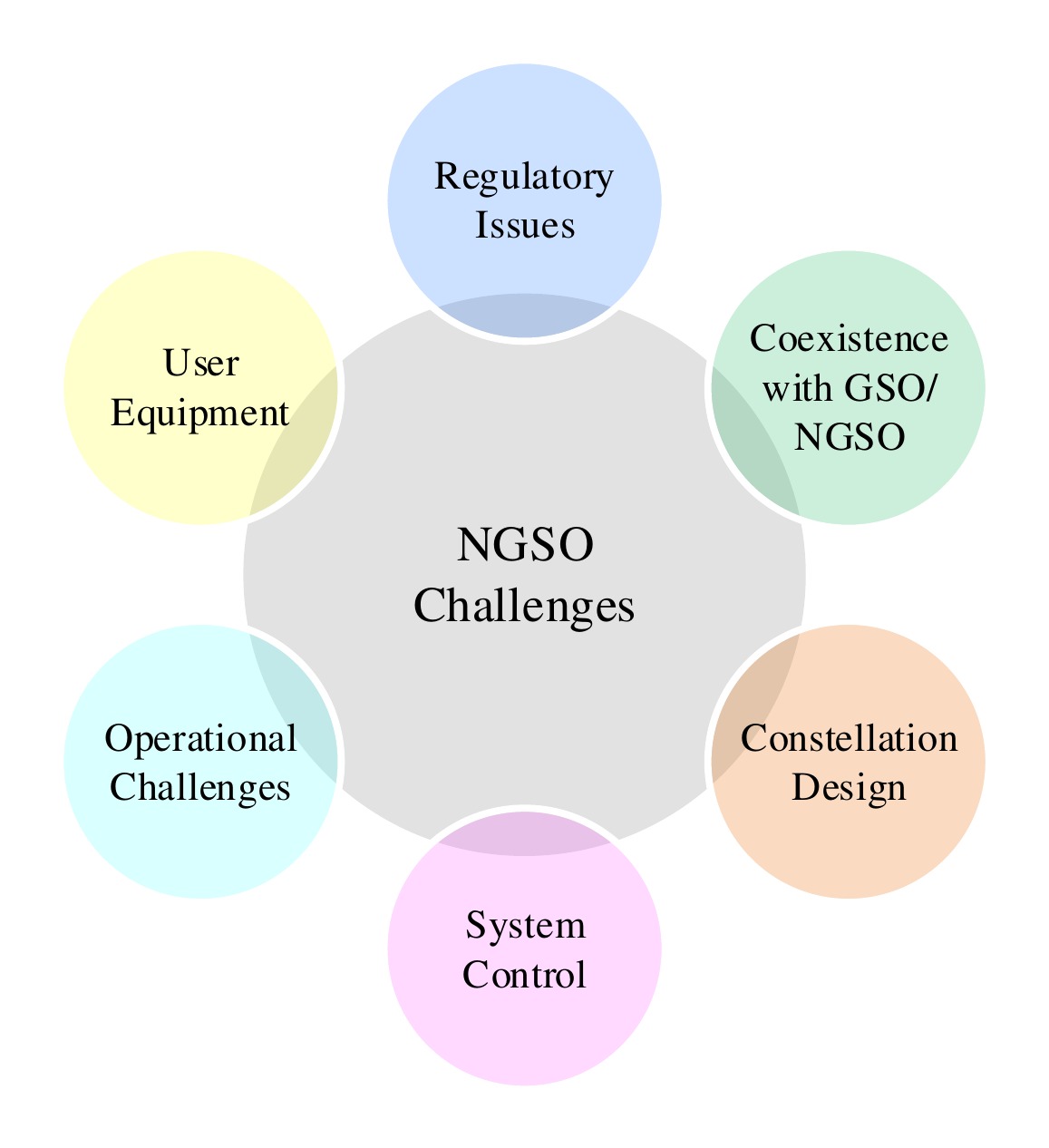}\vspace{-4mm}
	\caption{NGSO satellites deployment challenges.}
	\label{fig:NGSO_challenges}
\end{figure}

Beyond the NTN involvement and from NGSO deployment standpoints, this section presents several key challenges (see Fig. \ref{fig:NGSO_challenges}) including satellite constellation and architecture designs, coexistence with GSO and other NGSO systems in terms of spectrum access and regulatory issues, system operational issues, and user equipment requirements. In the following, the related critical challenges of NGSO systems deployment and integration are discussed with highlighting the most relevant solutions.

%---------------------------------------------
\subsection{Regulatory and coexistence issues}
%---------------------------------------------
According to the ITU regulations, the interference inflicted at GSO satellites from NGSO satellite systems shall not degrade GSO satellites performance and shall not claim protection from GSO systems in the fixed-satellite and broadcasting-satellite services \cite{ITU2003}. Specifically, the effective power flux density (EPFD) within the frequency bands that are allocated to GSO systems and at any point on the Earth's surface visible from the GSO satellite orbit shall not exceed the given predefined limits in the ITU regulations. Although NGSO systems have potentials of global coverage and high performance, many of their regulatory rules were coined nearly two decades ago based on the proposed technical characteristics of NGSO satellites at the time. This is very challenging from a spectral coexistence viewpoint, and it will require much more agile systems. Moreover, the deployment of NGSO satellites is undergoing a significant densification comparing to existing GSO systems, which is leading to unprecedented inter-satellite coexistence challenges. The high interference levels will not only result from the enormous number of operating satellites but also from the expected high heterogeneity of the NGSO systems \cite{Dai_2019}. Therefore, it is imperative to scrutinize the interference interactions between different GSO and NGSO systems to ensure consistent hybrid deployment landscape \cite{Riviere2019}.

The recent growing activities concerning the use of NGSO satellite constellations have propelled  the regulatory environment towards adapting and extending their rules to ensure a safe and efficient deployment of NGSO operations. International regulators have the difficult task to establish a fair and transparent competitive framework for all satellite broadband players while prioritising the socioeconomic growth.
Specifically, during the world radio communications conference in 2015 (WRC-15) \cite{WRC15}, different national delegates have expressed their concerns on the increasing number of requests submitted for NGSO satellite systems operating in the Fixed-Satellite Service (FSS) subject to the EPFD limits in Article 22 and to coordination under No. 9.7B of the Radio Regulations (RR). Furthermore, the global satellite coalition (GSC) during WRC-19 has agreed on defining a regulatory framework for NGSO satellites to operate in the Q/V bands \cite{WRC2019}. They also have planned a new agenda item for WRC-23 to further study a number of issues including technical considerations related to space-to-space links, which will be important for global NGSO and hybrid NGSO-GSO networks. Moreover, the ITU vision for the next WRC-23 aims at bring the satellite industry forward to work together with governments to ensure a global perspective on connectivity that also addresses national and regional requirements. 

At this point, some aspects  and scenarios need further investigations in this direction, which are enumerated and briefly described in the following. 
\begin{itemize}
    \item NGSO and GSO coexistence: NGSO single-entry power flux density (PFD) limits in certain  parts  of  the  frequency  range  10.7-30 GHz are included in Article 22 of the RR since 2000, with the main goal to protect GSO systems operating in the same frequency bands. 
    Later, the single-entry PFD limit was found to be not enough as the number of NGSO satellites was growing at a rapid pace. This led to the definition of EPFD that takes into account the aggregate of the emissions from all NGSO satellites. An example of multiple NGSO systems causing interference to a GSO receiver is shown in Fig. \ref{fig:Agg_Interf}. In this direction, a specific software tool has been made available for operators and regulators to check these limits for specific NGSO satellites \cite{ITU1503}. ESA has also launched a separate activity to build its own simulator \cite{ESA_CONSTELLATION}. Moreover, a feasible  solution can be proposed through constructing large discrimination angle and exclusion zones are typically considered to limit interference with GSO communications systems \cite{PatentHiggins}.
    
\begin{figure}[!t]\centering
	\includegraphics[width=0.38\textwidth]{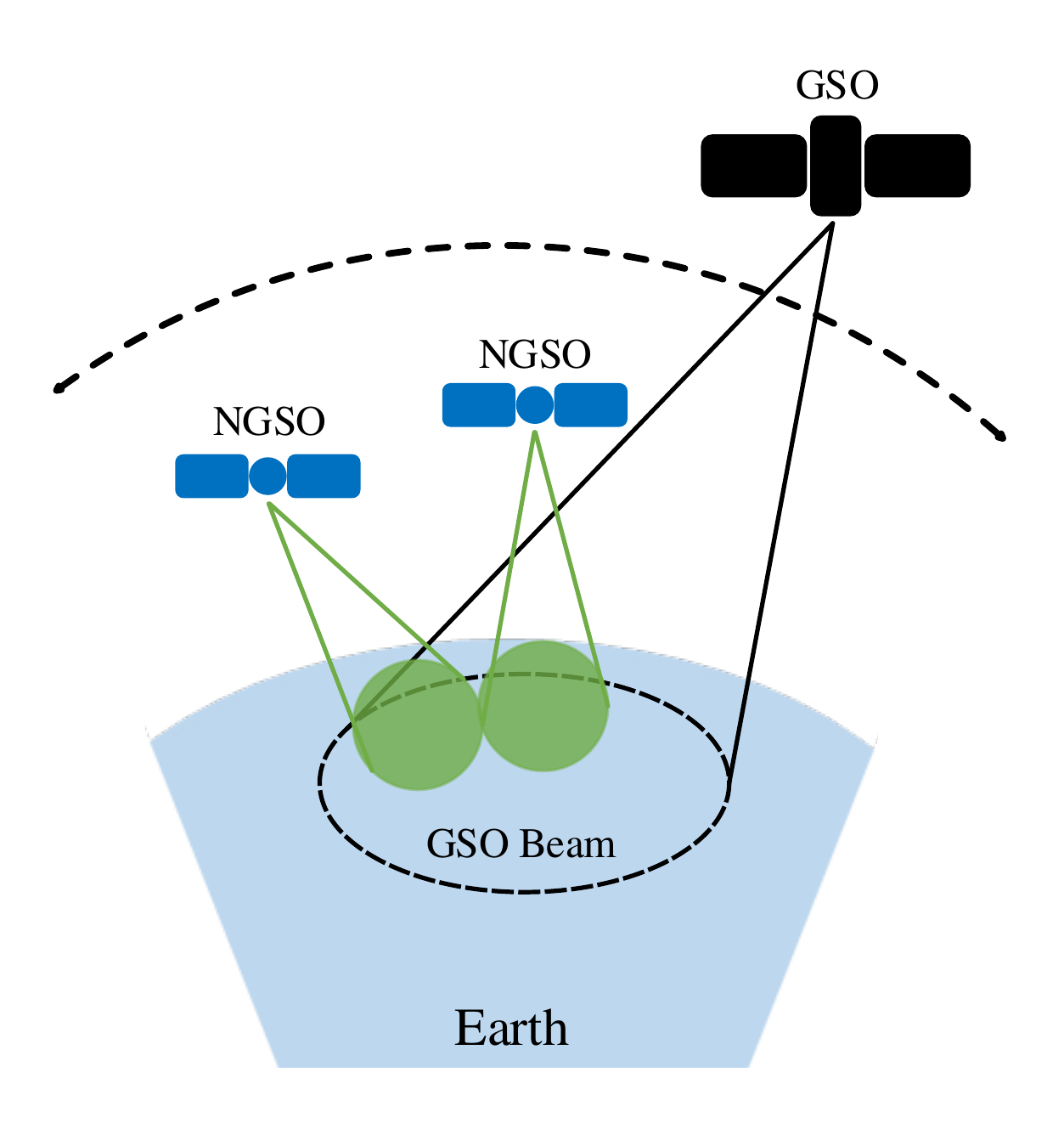}
	\caption{Aggregated interference from multiple NGSO systems.}
	\label{fig:Agg_Interf}
\end{figure}
    
    \item NGSO Earth stations operations: The ground infrastructure required to operate a NGSO constellation is significantly more complex than that of a single GSO satellite. Therefore, the impact of deploying multiple NGSO Earth stations distributed over the coverage area has to be carefully designed to ensure minimal impact on other users within the shared spectrum. However, from the regulators' perspective, there is no individual licensing of Earth stations because they believe that mitigation techniques can be employed by the operators to avoid detrimental interference, for example switching to alternative frequencies, as elaborated in Federal Communications Commission (FCC) documentations \cite{FCC17122}.

    \item NGSO FSS user terminals:  In general and excluding large latitudes, GSO FSS user terminals have significant gain in high elevation directions with limited gain towards the horizon, as the satellite is usually placed above the region of interest.
    Recently, advocates of a new generation of NGSO FSS systems have sought after the FCC authority to modernize the relevant regulations, and consequently, the FCC has proposed to update certain frequency allocations in the Ka-band, power limits, and service rules to facilitate these emerging systems \cite{FCC17122}. 
    
    \item Coordination with other NGSO networks: In view of the constellation and orbital overcrowding, it is very likely that large NGSO constellations will cause interference to other NGSO systems. However, the preliminary interference risk analysis carried out in \cite{Tonkin_2018} considering both Ka-band and V-band suggests that the risk is relatively low, concluding that the need for interference mitigation might be limited. In case of unacceptable interference situations, the mitigation  techniques  described  in  Annex  1  of  \cite{ITU1431}  should  be  considered  in order to  achieve satisfactory sharing between different NGSO systems, although other techniques are not excluded.
    
\end{itemize} 

It is clear that the efficient use of spectrum is one of the most crucial challenges to be met by international satellite community in order to mitigate the GSO-NGSO interference. While the NGSO inter-constellation interference is normally managed by ITU assigning priority based on the ITU filing date and without deteriorating the quality of service of GSO. The ITU regulation related to the NGSO-GSO spectrum sharing scenario is summarized in Table \ref{table:ITU_regulations} for the different bands of operations \cite{WRC2019}.

\begin{table}[!h] 
	\centering
	\caption{ITU regulations for NGSO-GSO spectrum sharing\cite{WRC2019}.} \label{table:ITU_regulations}
	\begin{tabular}{|l|l|l|} 
		\hline
		Band                 & Frequency Range    & Priority of Operations\\
		\hline
		Ku				&  \makecell{ 10.7-10.95 GHz (space-to-Earth)\\
			11.2-11.45 GHz (space-to-Earth)\\
			12.75-13.25 GHz (Earth-to-space)}   & \makecell{GSO has priority\\ over NGSO\\	EPFD limits apply}
	\\
		\hline
		Ka & \makecell{17.8-18.6 GHz (space-to-Earth)\\
			19.7-20.2 GHz (space-to-Earth)\\
			27.5-28.6 GHz (Earth-to-space)\\
			29.5-30 GHz (Earth-to-space)} & \makecell{GSO has priority\\ over NGSO\\	EPFD limits apply}\\
		\hline
		Q/V & \makecell{37.5-42.5 GHz (space-to-Earth)\\
			47.2-50.2 GHz (Earth-to-space)\\
			50.4-51.4 GHz (Earth-to-space)}  & \makecell{Maximum degradation\\ of GSO reference links:\\ \textbullet\ Single entry (3\%) \\ \textbullet\ Aggregate (10\%)
		}\\
		\hline
	\end{tabular}
\end{table}

\subsection{Satellite Constellation Design}
Generally, satellite orbit constellation design  is a key factor that directly affects the performance of the entire satellite systems. The fundamental constellation parameters include the type of orbit, altitude of the orbit, number of orbits, number of satellites in each orbit, and satellite phase factor between different orbit planes \cite{Qu2017}. Several earlier studies have considered systematic constellation patterns of satellites such as polar constellations and Walker-Delta patterns \cite{Walker1984}, which are formulated based on the relative positions of the satellites in the Earth-centered inertial (ECI) frame . Additionally, in \cite{Mortari2004}, the concept of flower constellations has been proposed to put all satellites in the same 3D trajectory in the Earth-centered Earth-fixed (ECEF) frame. However, these design approaches do not take into consideration the demand characteristics on Earth, which makes them inefficient  strategies when bearing in mind the non-uniform and uncertain demand over the globe. Thus, a more competent strategy would be a staged flexible deployment that adapts the system to the demand evolution and begins covering the regions that have high-anticipated demands.

Another relevant constellation concept that can be applied to NGSO systems was proposed in \cite{Paek2012} to constitute reconfigurable satellite constellations where satellites can change their orbital characteristics to adjust global and regional observation performance. This concept allows establishing flexible constellation for different areas of interest. However, introducing reconfigurability feature to the constellation requires a higher maneuvering capability of the satellites and more energy consumption and that can be a deterrent factor when multiple successive reconfigurations are needed over the life cycle. On the other hand, a hybrid constellation design is proposed in \cite{Chan2004} to utilize multiple layers and mixed circular-elliptical orbits, and thus, accommodating the asymmetry and heterogeneity of the traffic demand. Nonetheless, the optimization of adapting the constellation to growing demand areas is a challenging issue to be addressed in the context of integration an entire hybrid model. 
Moreover, an integrated framework that accounts for the spatial-temporal traffic distributions and optimizes the expected life cycle cost over multiple potential scenarios can be an initial plan to circumvent the NGSO constellation design challenges \cite{Ma2013}.

Furthermore, traditional global constellation systems are no longer valid solutions for NGSO systems due to high cost and inflexibility to react to uncertainties resulting from market demands and administrative issues. Therefore, regional coverage constellations are promising solutions for satellite operators as they will be able to tackle the economic and technical issues in a flexible manner \cite{Lee2020}. Regional constellations focus on the coverage over a certain geographical region by using a small number of satellites in the system and they can achieve the same or better performance compared to global-coverage constellations.
Regional coverage constellations can also provide sufficient redundancy with deploying  multiple NGSO satellites in lieu of a single GSO satellite, and thus, operators can hand off traffic to satellites that avoid beam overlapping, and therefore interference \cite{LEE2018213}.
However, designing an optimal regional constellation is a complicated process, which requires optimizing the orbital characteristics (e.g., altitude, inclination) while considering asymmetric constellation patterns, particularly for complex time-varying and spatially-varying coverage requirements. This topic has not been deeply investigated in the literature, and thus, new sophisticated approaches to design optimal constellation patterns are needed to be developed and  tailored to different orbital characteristics and NGSO environments.

%---------------------------------------------
\subsection{User Equipment}
%---------------------------------------------
Lowering latency of satellite communications can only be achieved by moving satellites closer to Earth, i.e., the low altitude NGSO satellites offer much lower latency compared with GSO. The closer a satellite is placed, the faster its movement is perceived from the user terminals on Earth, which imposes additional challenges to the user terminal equipment because it has to be able to track the satellite movement and perform handover from one satellite to another \cite{Jakoby2020}. The complexity of user equipment has an impact on its cost, which has been identified as a potential barrier for the commercial success of NGSO satellite communication systems. Previously, broadband LEO networks required expensive user equipment composed of mechanical gimbaled antennas, which has narrowed their roll out to only the customers with the high purchasing power mainly within the enterprise market \cite{Tang2021a}. Thus, a new generation of antenna and terminal technology was needed that should be affordable, easy to use, and adaptive to the increasingly complex space ecosystem. In other word, inexpensive user equipment capable of tracking LEO satellites are a significant component for widespread adoption and crucial to the business success of NGSO systems. 
In this context, AST \& Science initiative envisions building a space-based cellular broadband network to be accessible by standard smartphones where users will be able to automatically roam from land networks to a space network \cite{ast_science}. 

Conventional parabolic antennas provide good directivity at the expenses of costly mechanical steering \cite{Cheng_2012}. The continuous narrow beam pointing is a difficult task, which has pushed the ground equipment developers to fight in the battle of technical innovations. Electronic beam steering via antenna arrays, which have thus far been mainly used for military applications, are gaining momentum not only for NGSO satellites but also for moving platforms \cite{Guidotti2019a}. Low-cost and high-performance beam-tracking antennas are considered as a game-changer for the satellite community, and several companies are in the final stages of sending their products to the market, e.g., C-ComSat Inc, Kymeta, and ViaSat. Other antenna manufacturers are developing advanced silicon chips that can be used as building blocks of smart digital antennas to create electronic steered multi-beam array antenna \cite{Sadhu2019}. For instance, the startup Isotropic Systems has been working on developing modular antenna systems that are able to track more than one satellite at a time with a single antenna, which will enable multi-orbit operations and reduce the cost by combining their assets into a single integrated terminal without needing to duplicate circuity \cite{isotropic}.

Parabolic antennas are difficult to install, to configure and to operate, but they will still be dominant in governmental institutions and big moving platforms like cruise ships \cite{Choni2018}. Nevertheless, electronically steerable flat panel antennas are an imperative ground segment innovation offering a more agile, affordable and scalable antenna product capable of performing the same function as parabolic antennas, opening the door to the NGSO services to also small user terminals \cite{Tang2021a}. 
User mobility is another challenge to be addressed using inexpensive antennas. Interestingly, manufacturing a small, low-cost, flat-panel antenna that can be installed on various mobile assets seems feasible with employing the electrically steerable flat panel antennas. Moreover, ground equipment can benefit from satellites that have more flexibility and on-board processing capabilities that allow creating small and high power-beams over certain regions or assets, and that will change dramatically  how the landscape leverages the assets in the sky to facilitate user connectivity on ground \cite{Takahashi2019}.

Furthermore, the engagement of satellite industry with the 3GPP to integrate satellite networks into the 5G ecosystem yields an outcome that handheld users can be served by LEO and GSO in S-band with appropriate satellite beam layouts \cite{Sedin2020}. Besides, other users with high transmit and receive antenna gains (e.g., VSAT and proper phased array antenna) can be served by LEO and GSO in both S-band and Ka-band \cite{3GPP38821v16}. This also requires 5G functionalities to take into account the issues of long propagation delays, large Doppler shifts, and moving cells in NTN, and to improve timing and frequency synchronization. The characteristics of this user equipment are specified in \cite{3GPP38821v16}. In particular, the VSAT user equipment consists of a directional antenna (i.e., phased array antenna) with circular polarization and 60 cm equivalent aperture diameter, whereas the handheld user has an omnidirectional antenna element (e.g., dipole antenna) with linear polarization \cite{Dubovitskiy}.

%---------------------------------------------
\subsection{Operational Issues}
%---------------------------------------------
Other NGSO operational challenges/concerns are raised by the astronomy community as some rough estimates suggest there could be more than 50,000 satellites in total added to Earth orbits in the near future, which will make our planet blanketed with satellites. Therefore, some experts are alarmed by the plans of mega-constellation companies and raised many concerns specifically about the defunct satellites and smaller pieces of space debris \cite{Boley2021}. Additionally, astronomers have already expressed their disquiet about the resulting light pollution from the massive number of visible satellites, which will probably affect their scientific observations of the Universe. Thus, these concerns are briefly discussed next.

\begin{itemize}
    \item Light pollution: The proliferation of LEO satellites at altitudes less than 2,000 km will jeopardize the ability to observe, discover and analyze the cosmos from the Earth's surface. The astronomy community claims that the number of visible satellites will outnumber the visible stars and that their brightness in both optical and radio wavelengths will significantly influence their scientific research \cite{McDowell_2020}. A major issue with commercial satellite constellations is their visibility from the ground, where the prime contributing factor to light pollution from satellite constellations is the satellites’ size. However, currently there are a few mitigating options that can be considered to alleviate these concerns, which are presented in \cite{Walker2020Impact}. For instance, making satellites as small as possible, minimizing the reflectivity of  satellites, and providing the most accurate satellite orbits to understand observational ``avoidance zones'' by time or location for astronomy. The authors in \cite{Venkatesan2020} have called this issue an ``unfortunate irony'' because the technology indebted to centuries of study of orbits and electromagnetic radiation from space now holds the power to prevent the astronomical community from further exploration of the Universe. To this direction, the international astronomical research community has been active seeking a seat at decision-making tables to mitigate the impact of satellite mega-constellation on astronomical research.

    \item Space debris: Since the commercialization of NGSO satellites enters the realm of technical feasibility, many orbital debris concerns have been raised due to the long-term impact that results from placing thousands of satellites in orbits and the risk of causing  satellite collisions. Moreover, the advent large constellations of NGSO satellites have been added to the existing debate about the long-term impact of distributed spacecraft missions on orbital debris propagation. Thus, the field of studying the orbital debris is evolving in order to examine the potential debris mitigation strategies. For example, the work in \cite{Foreman2017} investigates the  impact of large satellite constellations on the orbital debris environment and uses OneWeb, SpaceX, and Boeing proposals as case studies. Authors in \cite{Kelly2019} study retrieving and relocating large debris for placement into the ``graveyard'' orbit above the geostationary regime as a way to mitigate orbital debris congestion. This work derives an analytical deorbit solution based on Lyapunov control theory combined with the calculus of variations. Another cost-effective way to diminish satellite debris has proposed to use a high power pulsed laser system on the Earth to make plasma jets on the objects, slowing them slightly, and causing them to re-enter and burn up in the atmosphere \cite{Phipps2012}. 
    
\end{itemize}

Information sharing and cooperation between regulators, astronomers, and industry could help facilitate the establishment of industry best practices and standards to ensure the long-term sustainability of both ground-based astronomy and satellite constellations. In this direction, the ITU radio astronomy recommendations are devised with policy protections for radio astronomy service (RAS) from interference by out-of-band emissions, e.g. as satellites broadcasting signals must redirect or cease such signals when passing over radio astronomy facilities. Additionally, groups like American Astronomical Society (AAS) and the International Astronomical Union (IAU) already act as representatives of the larger astronomy community, and they are actively expressing and discussing  astronomer's concerns about satellite constellations with regulators. The successful models that resulted in progress for other space sustainability issues like the United Nations working group on the ``long-term sustainability of space'' can be a good example to follow for regulators.

In a nutshell, in the landscape of future communications, employing NGSO satellites within the terrestrial networks and NTNs is seen as a cornerstone for accomplishing heterogeneous global communication systems with enhanced user experience. However, the success of the NGSO constellations hinges on several factors including a well-organized coexistence with other communication systems based on the regulatory procedures along with introducing optimal constellation patterns, and the developments of suitable user equipment.
%=======================================================
\section{Future Research Directions and \\ Opportunities} \label{sec:future_research}
%=======================================================

Evidently, NGSO satellites will be an essential part of our future communication systems, where they will converge with other wireless systems to achieve ubiquitous coverage, hybrid connectivity, and high capacity. Satellite technologies are under constant development to respond to the fast-changing demands of contemporary commercial and governmental systems through significantly higher capabilities and in a cost-effective manner. The disruptive potential of NGSO satellites does not lay only in serving the poorly connected areas but it also promises to open new frontiers for digital innovation. In this section, we present some futuristic visions and innovative research directions inspired by utilizing NGSO systems to further advance satellite communications within versatile applications.

%------------------------------------------------------------------------
\subsection{Open RAN}
%------------------------------------------------------------------------
Open Radio Access Network (ORAN) initiatives are developed to split  Radio Access Network (RAN) into multiple functional parts thereby enabling the interoperability of the vendor-independent off-the-shelf hardware and openness of software and  interfaces \cite{Singh2020a}.
Furthermore, the movement of ORAN actively promotes disaggregated RAN architectures enabled by standardized communication and control interfaces among the constituent components. The goal is to empower the innovation, enhance the security and increase the sustainability. The ORAN Alliance \cite{ORAN} actively promotes these initiatives. Furthermore, ORAN has found its way into 3GPP standardization \cite{ORAN2}.

All these aspects are very beneficial for satellite communication systems. For comparison, current satellite networks mostly rely on the implementation by a single manufacturer. This sole manufacturer usually provides all necessary network components, which are ``hard-wired'' within the system without any possibility to reconfigure. Hence, such vendor-dependent satellite networks lack flexibility and adaptability, especially for longer missions of more than 10 years because satellite hardware components can hardly be replaced. On the other hand, the persistent growth of the traffic demand and number of services with varying requirements, demand timely updates of the network configuration. In this context, ORAN offers the possibility to easily exchange the components with more advanced ones or extend the network by incorporating additional infrastructure \cite{papadias2020}. Thus, the advent ORAN architecture  is foreseen as a step towards a software oriented infrastructure that enables networks to operate based on the QoS requirement of the processed application.

For the emerging ORAN architecture, a novel strategy for the network management has been proposed in \cite{ORAN3,ORAN4}, which is based on AI and machine learning (ML)-driven policy definitions and resource management . This strategy enables the AI/ML-based solutions to the computationally intense tasks and the decision-making triggered by the network itself.
    
For NGSO satellite networks, the reconfiguration capability and vendor independence of ORAN are of special interest, since they allow a flexible extension of the constellation by adding more satellites or replacing their hardware and software with non-proprietary updates, which may work more efficiently in future. In this context, there are various challenges, since the compatibility of such diverse hardware may require a careful system design. In particular, the availability of data and the way how it is processed in different satellites needs to be taken into account. The most affected use cases for the application of ORAN seem to be resource management, carrier planning, and network adaptation. In addition, multi-layer mega-constellations seem to be the most demanding scenario for such an architecture. These use cases need to be analyzed in order to determine the price that needs to be paid for the enhanced flexibility of ORAN.

%------------------------------------------------------------------------
\subsection{Broadband Connectivity for Space Missions}
%------------------------------------------------------------------------
As discussed earlier, space-based Internet systems emerge as solutions to provide Internet access through a large number of LEO or MEO satellites. In addition to their unique capabilities in providing global coverage, low-latency communication, and high-speed Internet access points, they can dramatically change the way satellite missions are designed and operated in the near future. More specifically, the number of small satellite constellations in lower orbits for space downstream applications, such as Earth observation, remote sensing, and IoT collection, is constantly increasing. Currently, downstream mission operators heavily depend on a network of ground stations distributed across the globe for the purpose of downlinking data and controlling small satellites through telemetry and telecommand (TT\&C). Therefore, one of the key challenges for future space missions is providing a real-time uninterrupted connectivity, which is fairly infeasible in current satellite system infrastructure due to the magnitude and cost of the needed gateway network on ground. Even though some innovative concepts towards ground network sharing have recently appeared, such as Amazon AWS ground station \cite{AWS} and Microsoft Azure Orbital \cite{Azure}, the number and duration of ground access sessions are most of the times limited, preventing real-time mission operation and continuous high-throughput downstreaming data. 

Assuming a scenario where small satellites for downstream applications can directly access the Internet via a space-based Internet provider in a higher orbit, the small satellites can be constantly connected to the network without depending on a private or shared distributed network of ground stations \cite{SIN_2021}. This is certainly a game changer for the design and operation of future downstream satellite missions, since the communication link has to be pointing towards the sky instead of the Earth. This approach can be also replicated for the space-based Internet providers to enable a larger degree of connectivity in space network topologies. Further, this structure can lead to more inexpensive and sustainable space systems by reducing the number of required ground stations, while achieving real-time and reliable space communications. 

Employing the space-based Internet systems in this context can provide coordination of multiple constellations and awareness of the operational characteristics of each counterpart system. Additionally, space-based Internet systems will allow a satellite system to function strategically by transmitting TT\&C data between small satellite terminals and the NCC on the ground. However, the expected connectivity improvement will be achieved at the cost of higher complexity that is essential for load balancing between satellite links and for finding paths with the shortest end-to-end propagation delay, as well as tackling the dynamicity of the nodes (e.g. high relative speeds, frequent handovers), which are yet  unexplored areas in the literature.

%------------------------------------------------------------------------
\subsection{Edge Computing} 
%------------------------------------------------------------------------
One of the main challenges for the operation of satellites in general and especially NGSO satellites is rather low information processing capabilities of the on-board processors \cite{lovelly2014framework}. Consequently, complex processing tasks, such as online optimization of the resource allocation strategy, data processing for Earth observation applications, data aggregation for IoT, etc., can hardly be executed using a single satellite processor. Instead, the processing can be done in a distributed manner by pushing it from the central unity, e.g. GSO satellite, to the edge, e.g. NGSO satellites \cite{zhang2019satellite,wang2019game,9372909}. Besides that, computation offloading via NGSO satellites has been proposed in various works, e.g. \cite{9344666}. Moreover, edge computing has emerged as promising solution to alleviate the high latency issue by deploying processing and storage resource closer to users, especially for resource-hungry and delay-sensitive applications. Thus, integrating edge computing into NGSO networks can improve the performance of satellite networks by providing near-device processing capability. In this system, large amount of data generated by users can be processed through NGSO satellites instead of redirecting it to other servers, which will reduce network traffic load and the processing delay. While this application seems very promising, its practical limitations and requirements are not yet fully understood as it has started to attract the attention of researchers only in the last few years.
%One of the main challenges for the operation of satellites in general and especially NGSO satellites is rather low information processing capabilities of the on-board processors \cite{lovelly2014framework}. Consequently, complex processing tasks, such as online optimization of the resource allocation strategy, data processing for Earth observation applications, data aggregation for IoT, etc., can hardly be executed using a single satellite processor. Instead, the processing can be done in a distributed manner by pushing it from the central unity, e.g. GSO satellite, to the edge, e.g. NGSO satellites \cite{zhang2019satellite,wang2019game,9372909}. Besides that, computation offloading via NGSO satellites has been proposed in various works, e.g. \cite{9344666}. While this application seems very promising, its practical limitations and requirements are not yet fully understood as it has started to attract the attention of researchers only in the last few years.

%------------------------------------------------------------------------
\subsection{Space-based Cloud}
%------------------------------------------------------------------------
Far from the common use of satellites as relay devices, the space-based cloud concept has emerged as a promising and secured paradigm for data storage over NGSO satellites, particularly in the context of big data technologies and applications \cite{Jia2017}. The key advantage of space-based data storage is providing complete immunity from natural disasters occurring on Earth. Furthermore, utilizing NGSO satellites for data storage can offer more flexibility to some cloud networks that are designed to transfer data globally regardless the geographical boundaries and terrestrial obstacles \cite{Huang2018}. For instance, mega-corporations and large organizations that are located at different global sites can share big data through a space-based cloud and benefit from the faster transfer rate comparing to the traditional terrestrial cloud networks, especially for delay-sensitive services.
%Thereby, NGSO satellites could expand their scope of missions for more than only operating as relay devices for communication networks.

In this perspective, a startup company named Cloud Constellation is planning to establish a space-based data center platform SpaceBelt \cite{Spacebelt} that is offering secure data storage through LEO satellites and well-connected secure ground networks. In this infrastructure, the data-storage system is built upon multiple distributed satellites equipped with data-storage servers. However, the communication window between a ground station and an NGSO satellite is sporadic and the power budget in satellites is limited. Hence, this infrastructure imposes a significant challenge on developing scheduling algorithms for energy-efficient downloading files from the space-based data centers to meet dynamic demands of users under time-varying channel conditions. Besides, the existing operational algorithms for task scheduling in terrestrial cloud data centers are not applicable to the space-based cloud infrastructures \cite{Huang2020}. 

%------------------------------------------------------------------------
\subsection{IoT via NGSO Satellites}
%------------------------------------------------------------------------
The flexibility and scalability properties of NGSO satellites make their employment within the IoT ecosystem more appealing to shape novel architectures that uplift the interoperability among a plethora of applications and services \cite{Qu2017}. Thus, by exploiting the relatively short propagation distances of NGSO satellite constellations, IoT terminals can be designed to be small-sized, long-life, and low-power, which is ideal for the IoT operation. Moreover, the reduced OPEX and CAPEX of NGSO satellites comparing to GSO ones render them into good facilitators for the deployment of efficient IoT services over wide geographical areas \cite{Bacco2018}. Hence, these exceptional features of NGSO satellites can unleash the full potentials of IoT, and that will establish a universal network with billions of worldwide interconnected devices.

In this direction, the 3GPP organization in its release 17 \cite{RP-193235} has studied the necessary changes to support Narrow-Band IoT (NB-IoT) over satellites, including both GSO and NGSO systems. The objective here is to identify a set of features and adaptations enabling the operation of NB-IoT within NTN structure with a priority on satellite access.
In this context, some works have already started to adapt and evaluate these protocols under the NGSO system constraints specifically the relative satellite motion \cite{chougrani2021nb, charbit2020space, kodheli2018resource, kua2021}. Nevertheless, the progress is still in an early stage and more research efforts are required for a seamless integration, particularly in connecting NGSO satellites to mobile or stationary IoT devices and supporting ultra reliable low latency communications. 
% However, there are many technical challenges in connecting NGSO satellites to mobile or stationary devices, and this task particularly requires  a unified network vision on providing hybrid connectivity and prototyping satellite technology to support the advancement in machine-to-machine communications and IoT technologies.

%------------------------------------------------------------------------
\subsection{Caching Over NGSO Satellites}
%------------------------------------------------------------------------
Benefiting from the high-capacity backhaul links and ubiquitous coverage, NGSO satellites can help bring content closer to the end users, and thus, these satellites can be considered as an option for data caching. NGSO satellites also have the ability to multi-cast data and quickly update the cached content over different locations \cite{Liu2018}. Additionally, the symbiotic relationship between satellite and terrestrial telecommunication systems can be exploited to create a hybrid federated content delivery network, which will substantially ameliorate user experience \cite{Vu2018}. 
Therefore, integration of NGSO satellites into future Internet with enabling in-network caching makes traffic demands from users for the same content to be easily accommodated without multiple transmissions, and thereby, more spectral resources can be saved along with reducing transmission delay. Further, a promising strategy in this context is the combination of caching with edge computing over NGSO satellites, such that data processing, content analysis and caching are seamlessly integrated and harmonized \cite{Qiu2019}.
However, the time-varying network topology and limited on-board resources in NGSO satellites have to be taken into account when designing caching placement algorithms alongside with their fast convergence and low complexity.

%------------------------------------------------------------------------
\subsection{Aerial Platforms and NGSO Coordination}
%------------------------------------------------------------------------
Aerial platforms including unmanned aerial vehicles (UAVs) and HAPS are expected to play a crucial role in 6G wireless network development owing to due to their wider coverage footprints, strong LoS links, and flexibility of deployment compared to terrestrial networks \cite{Alfattani2021}. 
The use cases of low-cost unmanned aerial vehicles (UAVs) as flying mobile base-station are rapidly growing to expand wide-scale coverage range and improve wireless network capacity. Integrating terrestrial, airborne, and satellite networks into a single wireless system could provide comprehensive and efficient services. Moreover, UAVs and HAPS offer a high degree of mobility and a high chance for the LoS connectivity, which makes them perfect mobile relays for the satellite-terrestrial links \cite{lee2020integrating}. The use of NGSO and especially LEO satellites seems very promising due to a much smaller latency compared to GSO satellites, which is a necessary condition for the proper functioning and autonomous operation of the UAVs \cite{8741719}.

By introducing UAVs as part of the integrated space-air-ground system novel types of networks have been envisioned \cite{8741719}, such as UAV-aided cognitive satellite-terrestrial networks \cite{8611345}, cell-free satellite-UAV networks as part of future 6G systems \cite{9174846}, etc. Specifically, massive integrated networks are envisioned with multiple satellite orbits as part of NGSO mega-constellations, multiple UAVs and HAPS. Such networks pose many challenges for the coordination, navigation and synchronization. Some of the challenges have been investigated in \cite{8434289,8286975,yoo2021} for FSO, RF and hybrid signaling.
Further, the typical impairments to be considered in this context are high Doppler shift, pointing errors and outdated CSI. Another challenge is the topology control and multi-hop signal routing for such dynamic networks.

To summarize the key takeaway messages from this sections, NGSO satellites are highly anticipated to be an important player in reshaping various vertical applications and covering-up the flaws of current terrestrial communication systems. Specifically, some forward-looking views on NGSO constellations are explored in terms of seamless broadband connectivity across the globe with low latency and high service density. Further, amalgamating edge computing and caching technologies with NGSO networks will enable data storage and processing solutions that are faster, safer, and far more flexible than traditional offerings. Additionally, NGSO satellites can be used as reliable data storage to construct space-based clouds. Finally, connecting various NGSO satellites, aerial platforms, and IoT terminals with the terrestrial infrastructure to construct a multi-layer integrated system can support real-time communications, massive data transmission, and systematized information services.

%===========================================
\section{Conclusions}\label{sec:conclusions}
%===========================================
The deployment of NGSO satellites has been trending over the recent years owing to their less free space attenuation, low-profile antenna, small propagation delay, and the reduced orbital injection cost per satellite. The successful realization of NGSO communication systems is being achieved by the ongoing development of new technologies and the growing interest and investments, which have indeed pushed the satellite communication potentials towards higher bounds that need to be explored to support the rapid proliferation of various space-based applications and services. In addition, NGSO systems can be employed to support the terrestrial networks to overcome their limitations to match the rapid 5G ecosystem evolution though increasing the offered coverage and network capacity.

This survey presents the uprising technologies and research outlook in the realm of NGSO satellite communication systems along with the key technical challenges to integrate NGSO satellites into the global wireless communication platforms. 
Particularly, we conducted a detailed study on various communication aspects of NGSO satellites starting from the physical layer up to the applications and the entire structural design visions. Specifically, a detailed study of different physical connectivity and radio access schemes for multi-orbit satellites have been presented by reviewing the developments on inter-satellite connectivity, active antenna systems, waveform design, and link diversity and multiplexing. Next, the progress of establishing space information networking paradigms to cater for the unprecedented complexity and the scalability requirements is provided. The efforts on evolving NGSO satellites within current communication systems and architectures are also explored in terms of radio resource optimization, interference management, spectrum sharing, and security issues.

Moreover, in addition to studying the restrictions due to the coexistence with GSO systems, constellation design and resource management challenges, and user equipment requirements are explored as well. 
Afterwards, several innovative visions and research directions motivated by utilizing NGSO systems to deliver highly reliable and efficient global satellite communications for various applications are highlighted. Ultimately, this article covers the communication aspects and deployment challenges of NGSO satellites in the hope that it would trigger more in-depth investigations and serve as a continuous incentive for further NGSO research activities.

\linespread{1.09}
% ===========================================================================
% bibliography
% ===========================================================================

\bibliographystyle{IEEEtran}
\bibliography{IEEEabrv,References}

\end{document}